\documentclass[symmetry,review,accept,oneauthor,pdftex]{Definitions/mdpi} 

\usepackage{environ}
\NewEnviron{myequation}{%
\begin{equation}
\scalebox{1.3}{$\BODY$}
\end{equation}}

\firstpage{1} 
\pubvolume{1}
\issuenum{1}
\articlenumber{0}
\pubyear{2021}
\copyrightyear{2021}
\externaleditor{{Academic Editor: Firstname Lastname} 
} 
\datereceived{29 June 2021} 
\dateaccepted{27 July 2021} 
\datepublished{} 
\hreflink{https://doi.org/} 

\newcommand{\be}{\begin{equation}}
\newcommand{\ee}{\end{equation}}
\newcommand{\pa}{\partial}
\newcommand{\ts}{\tilde{\sigma}}
\newcommand{\nn}{\nonumber}
\newcommand{\unit}{\mathbb{I}}

\usepackage{graphicx}
\usepackage{slashed}
\usepackage{MnSymbol}
\pdfoutput=1

\Title{Degeneracy Patterns of Chiral Companions at Finite Temperature}

\TitleCitation{Degeneracy Patterns of Chiral Companions at Finite Temperature}

\Author{Juan M. Torres-Rincon \orcidA{}}

\AuthorNames{Firstname Lastname, Firstname Lastname and Firstname Lastname}

\AuthorCitation{Torres-Rincon, J.M.}

\address[1]{%
Institut f\"ur Theoretische Physik, Johann Wolfgang Goethe-Universit\"at, Max-von-Laue-Stra{\ss}e 1,\linebreak D-60438 Frankfurt am Main, Germany; torres-rincon@itp.uni-frankfurt.de}

\abstract{Chiral symmetry represents a fundamental concept lying at the core of particle and nuclear physics. Its spontaneous breaking in vacuum can be exploited to distinguish chiral hadronic partners, whose masses differ. In fact, the features of this breaking serve as guiding principles for the construction of effective approaches of QCD at low energies, e.g., the chiral perturbation theory, the linear sigma model, the (Polyakov)--Nambu--Jona-Lasinio model, etc.
At high temperatures/densities chiral symmetry can be restored bringing the chiral partners to be nearly degenerated in mass. At vanishing baryochemical potential, such restoration follows a smooth transition, and the chiral companions reach this degeneration above the transition temperature. In this work I review how different realizations of chiral partner degeneracy arise in different effective theories/models of QCD. I distinguish the cases where the chiral states are either fundamental degrees of freedom or (dynamically-generated) composed states. In particular, I discuss the intriguing case in which chiral symmetry restoration involves more than two chiral partners, recently addressed in the literature.}


\begin{document}

\section{Introduction}

Chiral symmetry is one of the key properties of the quantum chromodynamics (QCD), and it is one of the best principles for modeling the low-energy domain of the theory of strong interactions~\cite{Nambu:1960xd,Nambu:1961tp,Goldstone:1962es,Weinberg:1966fm,Glashow:1967rx,Weinberg:1967kj}. In vacuum, chiral symmetry is spontaneously broken and manifested in the absence of parity hadron doublets in Nature. In particular, the realization of its spontaneous symmetry breaking (SSB) is fundamental to constrain the internal structure of different effective approaches of QCD~\cite{GellMann:1960np,Nambu:1961tp,Weinberg:1966fm}. As an example, the pattern of the chiral group breaking for $N_f$ flavors, $SU_L(N_f) \times SU_R(N_f) \rightarrow SU_V (N_f)$, allows for the development of the chiral perturbation theory (ChPT) by incorporating the dynamics of the $N^2_f-1$ Goldstone bosons associated to the SSB~\cite{Gasser:1983yg}. Even when the chiral symmetry is already broken due to finite bare quark masses~\cite{Gasser:1982ap}, these can be incorporated in a consistent power counting so that the chiral expansion of the ChPT is well under control~\cite{Gasser:1983yg}.

The topic of chiral symmetry in the context of high-energy physics is very broad, and it has been studied in a number of works. For a broader perspective I refer the reader to the set of references~\cite{Pagels:1974se,Coleman:1985rnk,Hatsuda:1994pi,Leutwyler:1994fi,Bernard:1995dp,Alkofer:1995mv,Nowak:1996aj,Koch:1997ei,Cassing:1999es,Brown:2001nh,Hosaka:2001ux,Shuryak:2004pry,Bicudo:2009cr}. In this (mini-)review I focus on one of the consequences of chiral symmetry breaking: hadrons can be classified in chiral partners of opposite parity, with distinct masses. Classical examples (which will be addressed later) are the $\pi$ ($J^P$ = $0^-$) and the $\sigma/f_0(500)$ ($J^P=0^+)$ states or the $\rho$ ($J^P=1^-$) and $a_1$ ($J^P=1^+$) mesons.

While the vacuum state breaks chiral symmetry, finite temperature ($T$) and/or density ($n$) can bring chiral symmetry to a (partial) restoration following a phase transition. This transition from the broken (or Nambu-Goldstone) phase to the symmetric (or Wigner--Weyl) phase can be first-order, second-order or a crossover depending on microscopical parameters or external conditions~\cite{MeyerOrtmanns:1996ea}. The lowest-dimension order parameter of this transition is the quark condensate $\langle {\bar q} q\rangle (T,n)$, which takes small values (or zero) at high temperatures/densities. In QCD with physical quark masses, the chiral transition at vanishing net baryochemical potential is known to be a crossover~\cite{Aoki:2006we} at \mbox{$T_c$ $\simeq$ 155 MeV~\cite{Aoki:2006br,Bazavov:2011nk,Bhattacharya:2014ara,HotQCD:2018pds,Borsanyi:2020fev}. }Recent results from lattice-QCD calculations in the massless case ($O(4)$ universality class) point to a transition temperature of $T=132$ MeV~\cite{Ding:2019prx,Kaczmarek:2020sif}. When chiral symmetry is restored by the temperature/density one expects that the masses of chiral partners tend to become degenerate above this transition temperature (as long as no other mechanism makes the hadronic states to dissolve, like the deconfinement transition which eventually takes place in QCD~\cite{Shuryak:1978ij,Shuryak:1980tp,Meisinger:2001cq,PHENIX:2004vcz,STAR:2005gfr,Shuryak:2014zxa}). The details on how such a degeneracy is achieved have been largely debated in the past. 

An interesting perspective of the chiral partner degeneracy is given by the behavior of thermal correlation functions~\cite{Shuryak:1993kg}. Thermal QCD sum rules, as presented in Reference~\cite{Kapusta:1993hq}, allow for different realizations of chiral symmetry restoration in the vector and axial-vector correlation functions, where the involved states do not necessarily maintain a quasiparticle picture above $T_c$~\cite{Kapusta:1993hq,Koch:1997ei}. Among the different possibilities, the dropping mass scenario~\cite{DeTar:1987xb,Brown:1991kk} and a subsequent mass degeneracy between the $\rho$ and $a_1$ mesons is one of the most popular ones. Such picture is supported by the results of several phenomenological models like the linear sigma model (L$\sigma$M), the quark-meson model and the Nambu--Jona-Lasinio (NJL) model. A mass reduction of the vector meson with temperature/density is also supported by some phenomenological works which connect the shift of the thermal masses with experimental signatures, like the detection of dilepton (electron or muon) pairs~\cite{Pisarski:1981mq}. Thermal modification of the $\rho$-meson spectral function would bring measurable modifications in the dilepton spectrum at invariant masses below the $\rho$ pole mass~\cite{Li:1995qm,Rapp:1997fs}. In a baryon-rich medium, where chiral symmetry can be restored due to density effects~\cite{Cohen:1991nk}, this idea was used to analyze experimental results at low collisions energies, showing evidences of a medium-dependent vector meson mass~\cite{Rapp:1999ej}.

In this review I will examine several scenarios for the chiral-partner mass degeneracy at finite temperature. The classification is based on the structure/nature of the chiral partners. By that I refer either a state being a fundamental degree of freedom of the effective theory, i.e., represented by an explicit quantum field in the Lagrangian or a dynamically-generated state, which emerges via the solution of a (nonperturbative) few-body equation, in terms of other fundamental states. In both cases finite-temperature effects are key to understand the thermal evolution of their masses and decay widths and how they become degenerate (or not) when chiral transition temperature is approached. Whenever the (thermal) decay width is much smaller than the thermal mass, I will also denote the first class of states as {\it quasiparticles}; whereas the states which are only generated through attractive interactions, I will generically call {\it collective excitations}. A particular state---say, the pion $\pi$---could be of either type depending on the microscopic model used to describe it. I will consider the example of light scalar and pseudoscalar mesons ($\sigma/\pi$), but I will also show examples in the vector-axial vector sectors, genuine diquark states and finally the situation in the heavy-flavor sector. For that purpose, I will employ different effective theories/models at finite temperature.

Based on the previous classification, I will consider four distinct scenarios depending on the nature of the chiral partners:

\begin{enumerate}
\item Two fundamental degrees of freedom. The $\pi$ and $\sigma$ are explicit propagating fields in the effective Lagrangian. This case will be studied under the L$\sigma$M in the limit of a large number of boson fields. I address this case in Section~\ref{sec:lsm}.
\item Two composed (or generated) chiral states. The $\pi$ and $\sigma$ are mesonic excitations of the quark-antiquark attractive interaction. This is the case in the (Polyakov-)NJL model, which uses quark degrees of freedom with an effective interaction. This case is treated in Section~\ref{sec:pnjl}.
\item One fundamental and one composed state. Such is the situation of ChPT, where $\pi$ is a degree of freedom of the effective Lagrangian, while the $\sigma$ is a dynamically-generated state upon the unitarization of the $\pi-\pi$ scattering amplitude. This hybrid system is described in Section~\ref{sec:chpt}.
\item One fundamental and a two-pole composed state. This case incorporates an additional state to the study of chiral restoration. I will focus on the charm sector, where the $D$ meson is a fundamental degree of freedom of the effective Lagrangian, and the positive parity $D^*(2300)$ is a dynamically-generated state, which presents a two-pole structure. 
Section~\ref{sec:heavy} is devoted to this interesting case.
\end{enumerate}

The last case represents a novel situation which was considered at finite temperature for the first time in References~\cite{Montana:2020lfi,Montana:2020vjg}. In the present review I will address this system and speculate about how the chiral symmetry restoration can be reflected where three different masses evolve with temperature toward the chiral symmetry restoration point.

\section{Fundamental Chiral Partners: Linear Sigma Model in the Large-$N$ Limit~\label{sec:lsm}}

In the simplest scenario the effective Lagrangian incorporates explicitly the quantum fields, whose excitations become chiral companions. Then, both parity states are fundamental degrees of freedom. This case is probably the most familiar to the reader. Nevertheless, I will review it here because it helps to set some notation and to introduce finite-temperature techniques. It also clarifies how the chiral symmetry is recovered at finite temperature above $T_c$, for both cases with and without an explicit chiral-symmetry breaking term.  

One model belonging to this class is the linear sigma model (L$\sigma$M), where in its minimal version, $N$ pions and a $\sigma$ field are incorporated. Chiral symmetry is realized linearly in the boson fields, as the $\sigma$ appears explicitly (in the non-linear version of the model, the $\sigma$ field is integrated out and only the pions remain. The resulting Lagrangian coincides with the one of chiral perturbation theory at the lowest order~\cite{Weinberg:1978kz,Gasser:1983yg}). Some references describing different aspects of the L$\sigma$M are~\cite{Coleman:1974jh, Dobado:1994fd,Bochkarev:1995gi, Dobado:1997jx,Petropoulos:2004bt,Chakraborty:2010fr,Seel:2011ju}. See also Reference~\cite{Cortes:2016ecy} for recent applications of this model in the context of the chiral symmetry restoration. In this section I employ the L$\sigma$M in the approximation of a large number of pions $N$, as considered in References~\cite{Coleman:1974jh,Dobado:1997jx,Dobado:2009ek,Dobado:2012zf}. 

The Euclidean Lagrangian of the model is
\be 
\mathcal{L}_E = \frac{1}{2} \pa_{\mu} \Phi_i \pa^{\mu} \Phi_i - \overline{\mu}^2 \Phi_i \Phi_i + \frac{\lambda}{N} \left(\Phi_i \Phi_i  \right)^2 -\epsilon \Phi_{N+1} \ , \label{eq:LSMLag}
\ee
where $\Phi$ is a multiplet of $N+1$ scalar fields, conveniently parametrized as $\Phi_i =(\pi_a,\sigma)$ (with $a=1,\dots,N$). The quartic coupling satisfies $\lambda>0$, and $\overline{\mu}^2$ is a positive parameter forcing the SSB in vacuum. The last term of~(\ref{eq:LSMLag}) with finite $\epsilon$, explicitly breaks the $O(N+1)$ symmetry to $O(N)$.

The vacuum expectation value (VEV) of the field---assumed to be homogeneous---is chosen to be in the direction of the last component,
\be 
v^2(T=0) = \langle \Phi_i \Phi_i \rangle  (T=0) = \langle \sigma^2  \rangle (T=0) \ .
\ee

This VEV is to be matched with the physical value of the pion decay constant,
\be
v^2(T=0)= f^2_{\pi} = NF^2 \ , \label{eq:fpi}
\ee
where in the last step the explicit $N$ dependence has been extracted out.

Minimizing the effective action, one obtains
\be 
f_{\pi} \simeq \sqrt{\frac{N \overline{\mu}^2}{2 \lambda}} + \frac{\epsilon}{4 \overline{\mu}^2} = f_{\pi} (\epsilon=0) + \frac{\epsilon}{4 \overline{\mu}^2}
= \alpha f_{\pi} (\epsilon=0) \ , \label{eq:fpi2}
\ee
where for convenience I have defined $f_\pi^2 (\epsilon=0)=N\overline{\mu}^2/(2\lambda)$. This corresponds to the VEV of the $\sigma$ field in the absence of any explicit symmetry breaking term. I have also introduced $\alpha=1+N\epsilon/(8\lambda f^3_\pi (\epsilon=0))$, which serves as a multiplicative factor relating $f_\pi$ and $f_\pi (\epsilon=0)$ at $T=0$.

For completeness, the $N$-scaling of the different terms reads,
\be
\lambda \sim \mathcal{O} (1), \quad \overline{\mu}^2 \sim \mathcal{O} (1), \quad F^2 \sim \mathcal{O} (1), \quad f^2_{\pi} \sim \mathcal{O} (N), \quad \epsilon \sim \mathcal{O} (\sqrt{N}) \ . \label{eq:scaling}
\ee

Once the vacuum of the theory is fixed, one allows for fluctuations around the expectation values. Because $\langle \pi_a \rangle =0$ the pion fluctuation is simply denoted as $\pi_a$, while for the last component one writes $\sigma = f_{\pi} + \ts$.

The Lagrangian (up to an irrelevant constant) contains cubic and quartic interactions of the fluctuations,
\begin{align} 
 \mathcal{L}_E &= \frac{1}{2} \pa_{\mu} \pi_a \pa^{\mu} \pi_a+\frac{1}{2} \pa_{\mu} \ts \pa^{\mu} \ts - \overline{\mu}^2 \pi_a \pi_a - \overline{\mu}^2 \ts^2 
 + \frac{2 \lambda f^2_\pi}{N}  \left( \pi_a \pi_a + 3 \ts^2 \right) \nonumber \\ 
&+ \frac{\lambda}{N} \left[ (\pi_a \pi_a)^2  + 2 \pi_a  \pi_a \ts^2 + \ts^4 + 4 f_{\pi} \ts^3 +4f_{\pi} \pi_a \pi_a  \ts \  \right] \ . \label{eq:LafterSSB}
\end{align}
 
From Equation~(\ref{eq:LafterSSB}) it is possible to extract the tree-level masses~\cite{Dobado:2012zf},
\begin{align} 
m^2_{\ts} & = -2 \overline{\mu}^2 + \frac{12 \lambda}{N} f_\pi^2 = 4 \overline{\mu}^2+3 \frac{\epsilon}{f_{\pi}} \ , \label{eq:sigmamass} \\
m^2_{\pi} & = - 2 \overline{\mu}^2 + \frac{4 \lambda}{N} f_{\pi}^2 = \frac{\epsilon}{f_{\pi}} \ .  \label{eq:pimass}
\end{align}

Equations (\ref{eq:sigmamass}) and (\ref{eq:pimass}) show that for $\epsilon=0$, the $N$ pions are massless excitations in accordance to the Goldstone theorem~\cite{Goldstone:1961eq} [$N=\dim O(N+1) - \dim O(N)$], while the $\sigma$-excitation receives a nonzero mass. On the other hand, for physical quark masses $\epsilon \neq 0$, and the pions do not remain massless anymore. In this approximation Equation~(\ref{eq:pimass}) can be used to fix $\epsilon$ in accordance to the physical pion mass, and then Equation~(\ref{eq:sigmamass}) can be employed to fix the parameter $\overline{\mu}^2$ inserting the vacuum mass of the $\ts$ excitation.

At finite temperature ($T\neq 0$) one takes advantage of the large-$N$ limit and computes the partition function,
\be 
\mathcal{Z} = \int \mathcal{D} \pi_a \mathcal{D} \sigma \ \exp \left( -\int d^4 x \mathcal{L}_E \right) 
\ee
by introducing and auxiliar field $\chi$ to perform the Gaussian integration.
The auxiliar field reads~\cite{Dobado:2012zf}
\be
\chi = 2 \sqrt{2} \frac{\lambda}{N} \Phi_i \Phi_i \ , 
\ee
so that
\be 
\exp \left( \int d^4x \frac{\lambda}{N} (\Phi_i \Phi_i)^2 \right) = \int \mathcal{D} \chi \exp \left[ - \frac{1}{2} \int d^4x \left( \frac{N}{4 \lambda} \chi^2 - \sqrt{2} \chi \Phi_i \Phi_i \right) \right] \ , 
\ee
up to an overall irrelevant constant.

The partition function reads
\begin{align} 
\mathcal{Z} & =  \int \mathcal{D} \pi_a \mathcal{D} \sigma \mathcal{D} \chi \  \exp \left\{ -\int d^4 x \left[
\frac{1}{2} \pa_{\mu} \pi_a \pa^{\mu} \pi_a + \frac{1}{2} \pa_{\mu} \sigma \pa^{\mu} \sigma - \overline{\mu}^2 \pi_a \pi_a - \overline{\mu}^2 \sigma^2 \right. \right.  \nn \\
& -  \left. \left.  \frac{1}{2} \frac{N}{4\lambda} \chi^2 + \frac{1}{2} \sqrt{2} \chi \pi_a \pi_a + \frac{1}{2} \sqrt{2} \chi \sigma^2 - \epsilon \sigma \right] \right\} \ .
\end{align}

As for the $T=0$ case, one first finds the minimum of the effective action and incorporate small fluctuations around it. After performing a shift to the $\sigma$ field, $\sigma=v+\ts$, the auxiliar field also receives an induced shift, $\chi=2\sqrt{2} \lambda v^2/N+\tilde{\chi}+4\sqrt{2} \lambda v\ts/N$~\cite{Dobado:2012zf}. The Euclidean action reads
\begin{align}
 S_E [v , \chi ; \pi_a,\ts] & = \int d^4x \ \left[  \frac12 \pi_a \ (-\square_E + G_\pi^{-1} [0,\chi] ) \ \pi_a + \frac12 \tilde{\sigma} \ (-\square_E+G_{\tilde{\sigma}}^{-1} [0,\chi]) \ \tilde{\sigma}   \right.  \nn \\
 &+ \left. \frac12 v^2 G_\pi^{-1} [0,\chi] - \frac{f_\pi^2}{2\alpha^2} G_\pi^{-1} [0,\chi] 
 - \frac{N}{16 \lambda} (G_\pi^{-1} [0,\chi])^2  - \epsilon v   \right] \ .
\end{align}
where I have introduced the functions,
\begin{align}
G^{-1}_{\pi} [q,\chi] & \equiv  q^2 - 2\overline{\mu}^2 + \sqrt{2} \chi \ , \label{eq:chiG} \\
G^{-1}_{\ts} [q,\chi] & \equiv  G_\pi^{-1} [q,\chi] + 8 \frac{\lambda}{N}v^2 \ .
\end{align}

These play the role of the inverse pion/sigma propagators at tree level. Notice that their representation in Fourier space is anticipated in the $q^2=|{\bf q}|^2$ term. While $\pi_a$ and $\ts$ are dynamical fields, $\chi$ has no kinetic term. At mean-field level, one can trade $\chi$ by the variable $G^{-1}_\pi [0,\chi]$ (they are linearly related by Equation~(\ref{eq:chiG})). The Euclidean action also depends on $v$, which is yet to be determined as a function of $T$ (still assumed to be homogeneous for simplicity).

One can integrate out the fluctuations $\pi_a$ ad $\ts$, leaving an effective action for the functions $G^{-1}_\pi [0,\chi]$ and $v$. Formally this step is understood as
\be  
\exp \left( -S_E^{{\rm eff}}[v,G^{-1}_{\pi} [0,\chi]] \right) = \int \mathcal{D} \pi_a \mathcal{D} \ts \exp \left( -S_E [v,G^{-1}_{\pi} [q,\chi];\pi_a,\ts \right) \ ,
\ee
where the integration is performed at finite temperature using the imaginary time formalism~\cite{Kapusta:2006pm}. In particular one finds,

\begin{align} 
& \int \mathcal{D} \pi_a  \exp \left(- \int d^4x \ \frac{1}{2} \pi_a \left[ - \square_E + G_\pi^{-1} [0,\chi] \right] \pi_a \right)  \rightarrow
 \int d^4x \exp \left( - \frac{N}{2} \sumint_q \log G_\pi^{-1} [q,\chi] \right) \ ,  
\end{align}
and similarly for the $\ts$ field. The symbol
\be 
\sumint_q \equiv T \sum_n \int \frac{d^3q}{(2\pi)^3} \ , \label{eq:sumint}
\ee
represents the sum over Matsubara bosonic frequencies ($i\omega_n=2iTn$ with $n\in \mathbb{N}$) and the integration over three-momentum.

The resulting effective potential density reads~\cite{Dobado:2009ek,Dobado:2012zf}
\begin{align}
 V_{ {\rm eff}} (v,G_\pi^{-1}[0,\chi];T) &= \frac{1}{2} \left( v^2 - \frac{f_\pi^2}{\alpha^2} \right) G_\pi^{-1} [0,\chi] - \frac{N}{16\lambda} (G_\pi^{-1} [0,\chi] )^2 - \epsilon v   \nn \\ 
 &+  \frac{N}{2} \sumint_q \log  G_\pi^{-1} [q,\chi] + \frac{1}{2} \sumint_q \log G_{\tilde{\sigma}}^{-1} [q,\chi] \ ,  \label{eq:Veff}
\end{align}
where I stress that $f_\pi$ is the vacuum value of $v$, for the physical pion mass ($\epsilon \neq 0$ case), cf. Equations~(\ref{eq:fpi}) and (\ref{eq:fpi2}).

The large-$N$ limit is invoked here to neglect the last term of Equation~(\ref{eq:Veff}). This can be understood by using the scaling properties of the different quantities given in Equation~(\ref{eq:scaling}). This approximation suppresses the effects of the $\sigma$-mode fluctuations, which in turn, should become relevant close to the chiral phase transition~\cite{Rajagopal:1992qz}. Therefore, this limit is basically equivalent to a mean-field approximation. To next order, the description around the phase transition should incorporate the fluctuation of the $\sigma$ field~\cite{Rajagopal:1992qz,Stephanov:1998dy,Stephanov:2008qz,Grossi:2021gqi}, and for instance critical exponents will take values beyond the mean-field predictions. For the purpose of studying the mass degeneracy above $T_c$, the leading order of the large-$N$ expansion is enough.

Finally one needs to renormalize the effective potential. I skip the technical steps and refer to References~\cite{Dobado:2009ek,Dobado:2012zf} for details. These are not very illuminating for the purpose of this work. The final form of the effective potential reads,
\begin{align}
 V_{ {\rm eff} } (v,G_\pi^{-1}[0,\chi] ; T) &= \frac{1}{2} \left(  v^2 - \frac{f_\pi^2}{\alpha^2} \right) G_\pi^{-1} [0,\chi] - \epsilon v - \frac{N}{2} g_0 (T, G_\pi^{-1}[0,\chi])  \nn \\ 
 &-  \frac{N}{16} (G_\pi^{-1} [0,\chi])^2 \left[ \frac{1}{\lambda_R(\mu)} - \frac{1}{4\pi^2} \log \left( \frac{e^{1/2} G_\pi^{-1} [0,\chi]}{\mu^2} \right) \right] \ , \label{eq:Veffren}
\end{align}
where the function $g_0(T,M^2)$ (finite thermal part of the Matsubara sum in Equation~(\ref{eq:Veff})) is defined as
\be
g_0(T,M^2) = \frac{T^4}{3\pi^2} \int_{M/T}^\infty dx \ \left( x^2- M^2/T^2 \right)^2 \ \frac{1}{e^x -1} \ , 
\ee
and $\lambda_R(\mu)$ is the renormalized coupling which depends on a regularization scale $\mu$. However, the effective potential does not depend on $\mu$ because the running of $\lambda_R$ cancels the explicit $\mu$ dependence in the last term of Equation~(\ref{eq:Veffren}).

The effective potential at finite temperature~(\ref{eq:Veff}) can be now minimized with respect to $v$ and $G_\pi^{-1}[0,\chi]$. These two conditions are known as ``gap equations'',
\begin{align} 
 \left.  \frac{\partial V_{ {\rm eff} } (v,G_\pi^{-1}[0,\chi]) }{ \partial v } \right|_{v=\overline{v}, G_\pi^{-1}[0,\chi]=\overline{G_\pi^{-1}[0,\chi]} } & = 0 \ , \label{eq:gap1} \\
 \left. \frac{\partial V_{ {\rm eff} } (v,G_\pi^{-1}[0,\chi]) }{ \partial G^{-1}_\pi [0,\chi] } \right|_{v=\overline{v}, G_\pi^{-1}[0,\chi]=\overline{G_\pi^{-1}[0,\chi]} } & = 0 \ , \label{eq:gap2}
 \end{align}
and they provide the thermal expectation value of $v$ (denoted $\overline{v}$) and $G^{-1}_\pi [0,\chi]$ (denoted $\overline{G^{-1}_\pi [0,\chi]}$). Once these equations are solved, the pion and the sigma thermal masses simply read~\cite{Dobado:2012zf}
\begin{align} 
m_\pi^2 (T) &= \overline{G^{-1}_\pi [0,\chi]}  \ , \label{eq:mpiT} \\
m_\sigma^2 (T)&= \overline{G^{-1}_\pi [0,\chi]} +8 \frac{\lambda_R}{N} \overline{v}^2 \ . \label{eq:msigmaT}
\end{align}

\begin{figure}[H]
\vspace{2pt}
\begin{center}
\includegraphics[width=0.35\textwidth]{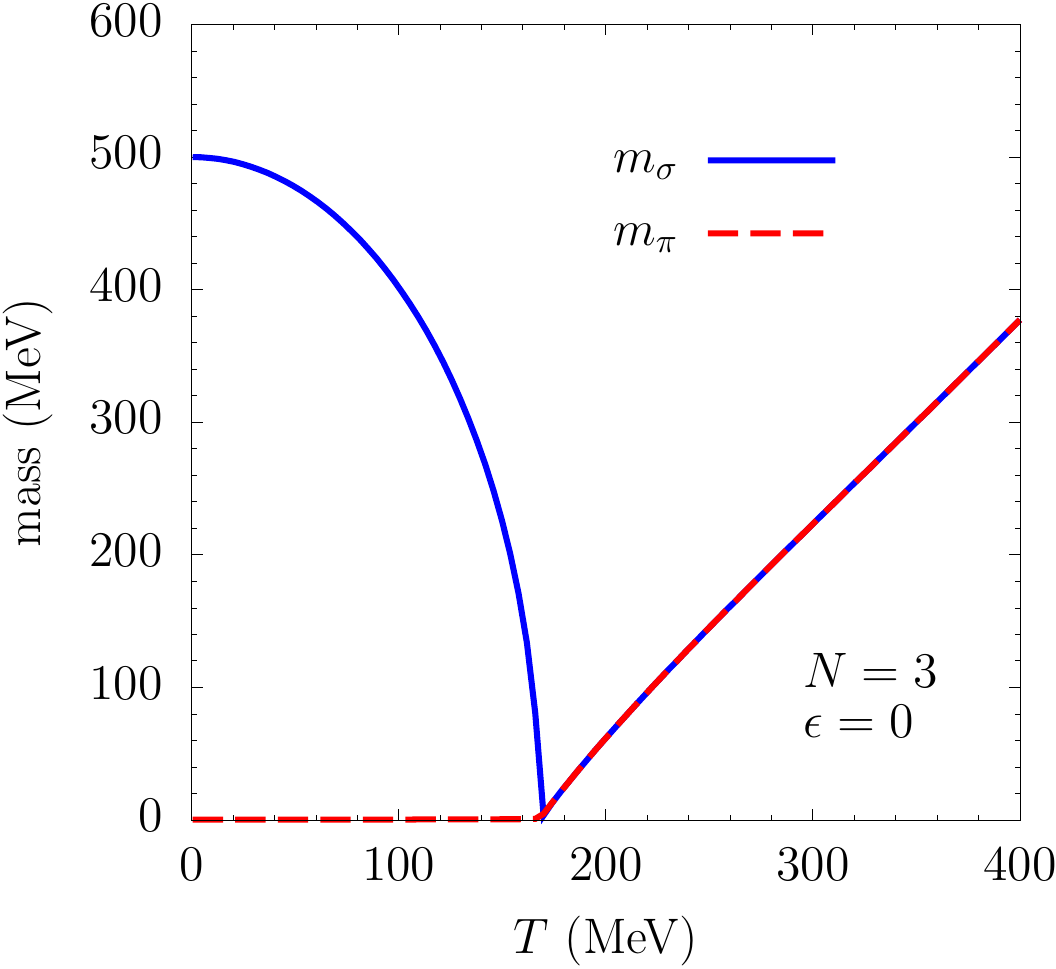}
\includegraphics[width=0.35\textwidth]{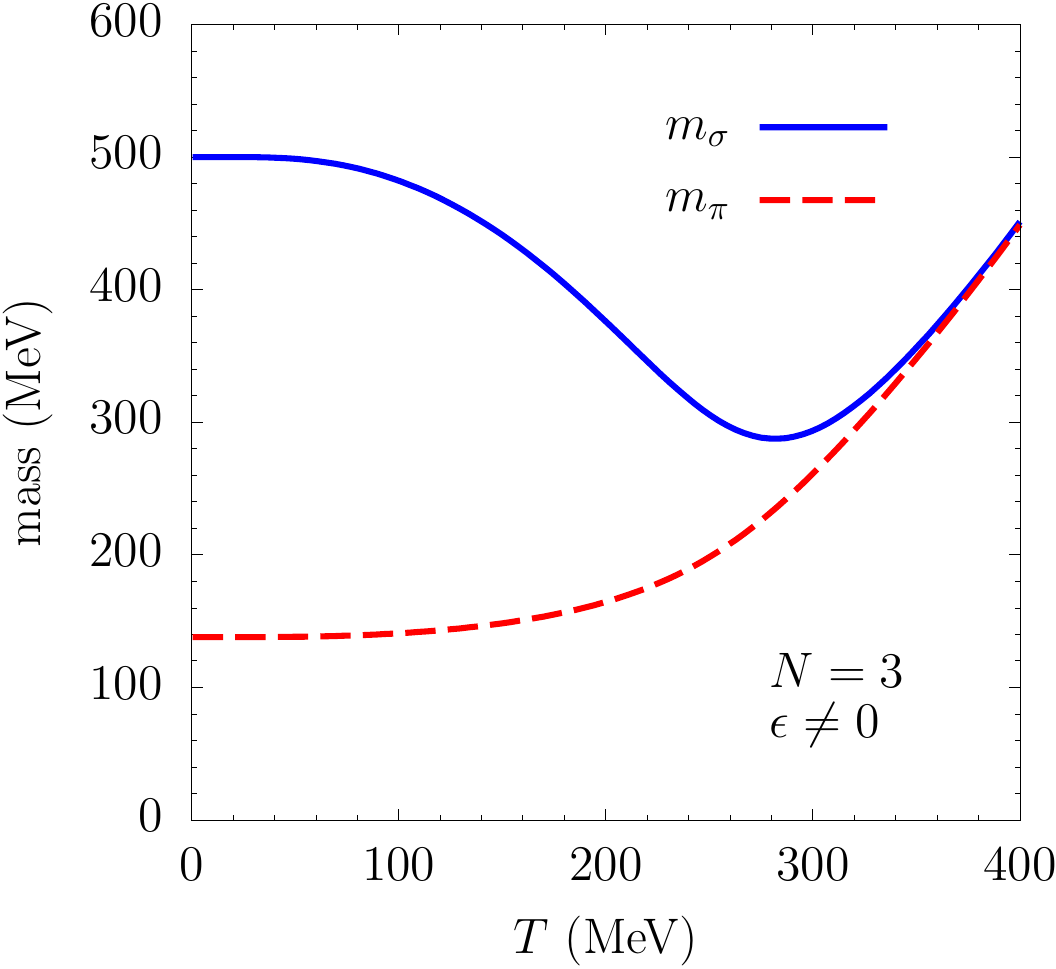}
\caption{$\pi$ and $\sigma$ masses as functions of the temperature below and above the chiral phase transition. (\textbf{Left panel}): Case without explicit symmetry breaking ($\epsilon=0$), where the pion is a true Goldstone boson for $T<T_c=168$ MeV. (\textbf{Right panel}): Case with explicit symmetry breaking ($\epsilon \neq 0$), where the pion is a pseudo-Goldstone boson, with a vacuum mass of 138 MeV. In both cases the pion and the $\sigma$ fields become degenerate at high temperatures.\label{fig:LSM}}
\end{center}
\end{figure}

I plot the $\pi$ and $\sigma$ masses in Figure~\ref{fig:LSM} for the cases $\epsilon=0$ (left panel) and $\epsilon \neq 0$ (right panel). In both situations the $\sigma$ vacuum mass is chosen to be $m_\sigma(T=0)=500$ MeV, and the pion mass (for the case $\epsilon \neq 0$) $m_\pi(T=0)$ =138 MeV.

The left panel of Figure~\ref{fig:LSM} presents a genuine SSB where $N$ Goldstone bosons are generated at low temperatures. For illustration, a value of $N=3$ is used in the equations. In the Nambu-Goldstone phase the renormalized mass of the $\sigma$ is fixed by the value of $\lambda_R$ parameter. As shown in Equation~(\ref{eq:msigmaT}), the temperature dependence of $m_\sigma$ follows closely the one of the order parameter $\overline{v}(T)$ and becomes zero at~\cite{Dobado:2012zf} $T_c=2\sqrt{3} F = 2 f_\pi (\epsilon=0)$, which in this case is $T_c \simeq 168$ MeV (I choose $\overline{\mu}^2$ so that $f_\pi=93$ MeV; therefore, \mbox{$f_\pi (\epsilon=0)=f_\pi/\alpha \simeq 84$~MeV}). In the chirally-restored phase (Wigner--Weyl phase) all the components of the $\Phi$ multiplet become degenerate, and the $\pi-\sigma$ chiral partners share the same mass. This can already be seen from Equations~(\ref{eq:mpiT}) and (\ref{eq:msigmaT}) by setting $\overline{v}=0$.

\textls[-25]{In the right panel of Figure~\ref{fig:LSM} I present the physical case, with an explici}t \mbox{$O(N+1)$-breaking} term. The three parameters ($\overline{\mu}^2, \lambda_R,\epsilon$) are fixed by setting the values of $(m_\pi,f_\pi,m_\sigma)$ in vacuum.  In this case both pion and $\sigma$ masses are nonzero at $T=0$. For increasing temperatures the masses approach each other, and at the phase transition---which happens smoothly, i.e., a crossover---they become close and the chiral symmetry has been \mbox{partially restored.}

\begin{figure}[H]
\begin{center}
\includegraphics[width=0.35\textwidth]{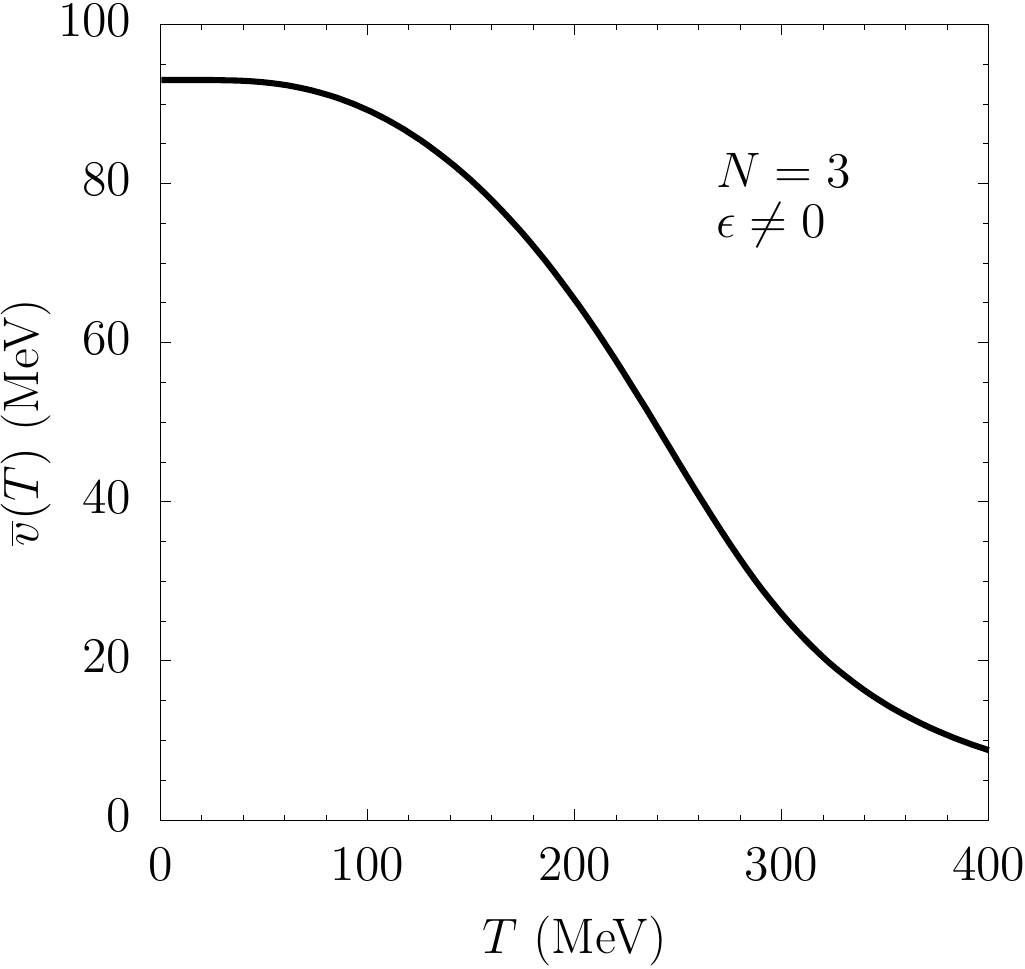}
\includegraphics[width=0.35\textwidth]{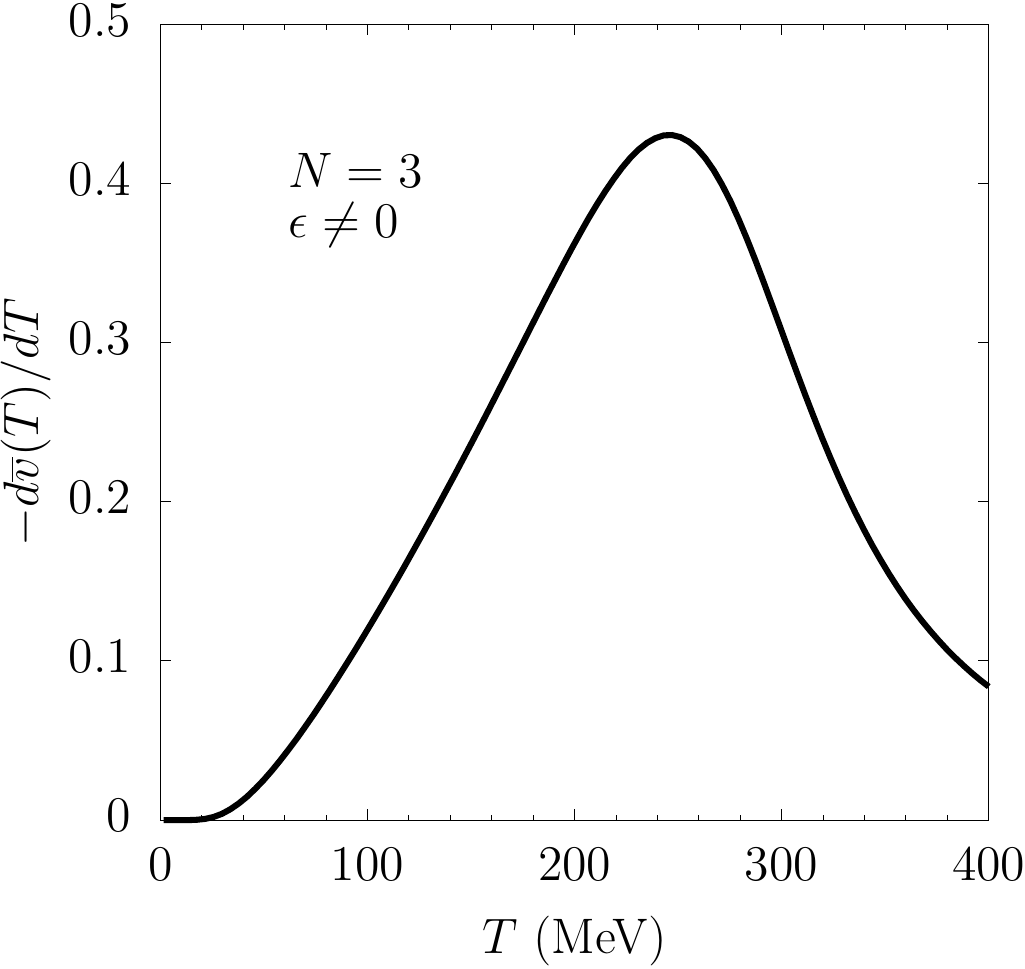}
\caption{\label{fig:vLSM} (\textbf{Left panel}): Order parameter of the chiral transition in the L$\sigma$M in the $\epsilon \neq 0$ case, corresponding to a crossover. (\textbf{Right panel}): Minus the $T$-derivative of the order parameter as a function of temperature. A measure of the location of the crossover transition is given by the maximum of \mbox{this quantity}.}
\end{center}
\end{figure}

The determination of $T_c$ is illustrated in Figure~\ref{fig:vLSM}. In the left panel I show the temperature dependence of the order parameter $\overline{v}$ obtained from the gap \mbox{Equations~(\ref{eq:gap1}) and (\ref{eq:gap2})} for the $\epsilon \neq 0$ case. The value of $\overline{v}$ at $T=0$ is the value $f_\pi=93$ MeV. Notice that the order parameter has an inflection point at some intermediate temperature, which can be used to determine the value of $T_c$. In the right panel I present (minus) the derivative of the order parameter with respect to temperature. The maximum of this quantity provides the value $T_c \simeq 246$ MeV. It is important to note that at this $T_c$ the $\pi$ and $\sigma$ masses are not yet degenerated, as can be seen in the right panel of Figure~\ref{fig:LSM}, and an approximate degeneracy only happens well above $T_c$.

To summarize the fist case described in this work, the L$\sigma$M clearly illustrates the thermal evolution of chiral partners ($\pi$ and $\sigma$) and the eventual mass degeneration once the chiral symmetry is restored for $T>T_c$. Both hadronic states appear in the original Lagrangian of the model and the thermal evolution of their masses is calculated by solving the gap Equations~(\ref{eq:gap1}) and (\ref{eq:gap2}), which describe how thermal fluctuations modify the propagation of the different excitations at a mean-field level. Mass degeneracy is clearly observed in this example.

\section{Composed Chiral Partners: (Polyakov--)Nambu--Jona-Lasinio Model~\label{sec:pnjl}}

In this section I will describe one of the best-known scenarios in which the chiral partners are not contained among the degrees of freedom of the Lagrangian. I will also consider the prototypical $\pi/\sigma$ case. However---as opposed to the previous example---both states will emerge as bounds states/resonances after applying a two-body equation.

One of the effective models able to describe chiral phenomena of QCD in the low energy regime is the Nambu--Jona-Lasinio (NJL) model~\cite{Nambu:1961tp,Vogl:1991qt,Klevansky:1992qe,Hatsuda:1994pi,Alkofer:1995mv,Buballa:2003qv}. It describes quarks at low energies where gluons have been integrated out. Some gluonic properties (like the expectation value of the Polyakov loop~\cite{Meisinger:2001cq}) can still be described by using the extension to the Polyakov--Nambu--Jona-Lasinio (PNJL) model~\cite{Fukushima:2003fw,Megias:2004hj,Ratti:2005jh,Hansen:2006ee,Fukushima:2008wg,Torres-Rincon:2015rma}. The description of the PNJL model given here is based on the conventions and definitions in References~\cite{Torres-Rincon:2015rma,Torres-Rincon:2017zbr}.

The Lagrangian density of the PNJL model with $N_f=3$  flavors is
\begin{align} 
{\cal L}_{PNJL} &= \sum\limits_i \bar{\psi}_i (i \slashed{D}-m_{0i}+\mu_{i} \gamma_0) \psi_i  \nn\\
&+ G \sum\limits_{a} \sum\limits_{ijkl} \left[ (\bar{\psi}_i \ i\gamma_5 \tau^{a}_{ij} \psi_j) \ 
(\bar{\psi}_k \ i \gamma_5 \tau^{a}_{kl} \psi_l)
+ (\bar{\psi}_i \tau^{a}_{ij} \psi_j) \ 
(\bar{\psi}_k  \tau^{a}_{kl} \psi_l) \right] \nn \\
&+ G_V \sum\limits_{a} \sum\limits_{ijkl} \left[ (\bar{\psi}_i \ i\gamma_5 \gamma_\mu \tau^{a}_{ij} \psi_j) \ 
(\bar{\psi}_k \ i \gamma_5 \gamma^\mu \tau^{a}_{kl} \psi_l)
+ (\bar{\psi}_i \tau^{a}_{ij} \gamma_\mu  \psi_j) \ 
(\bar{\psi}_k  \tau^{a}_{kl} \gamma^\mu \psi_l) \right]  \nn\\
& -    H \det_{ij} \left[ \bar{\psi}_i \ ( \unit - \gamma_5 ) \psi_j \right] - H \det_{ij} \left[ \bar{\psi}_i \ ( \unit + \gamma_5 ) \psi_j \right]  \nn \\ 
&- {\cal U} (T;\Phi,\bar{\Phi})\ , \label{eq:lagPNJL} 
\end{align}
where the quark field is labeled with flavor indices $i,j,k,l=\{u,d,s\}$, and $\tau^{a}$ ($a=1,\dots,8$) are the flavor generators of $SU_f(3)$ algebra, normalized as $\textrm{tr}_f \  (\tau^{a} \tau^{b}) = 2\delta^{ab}$. The covariant derivative is $D^\mu=\partial^\mu-i\delta^{\mu 0}A^0$, with $A^0$ the temporal component of the gluon field. $\mu_i$ are the possible chemical potentials for different flavors, which will be set to zero in this work. $G$ and $G_V$ are the scalar and vector couplings. $H$ is the 't Hooft six-quark coupling, and ${\cal U}$ is the Polyakov loop effective potential, parametrized for the Polyakov loop $\Phi$ and its conjugate $\bar{\Phi}$ as a function of the temperature (see \cite{Torres-Rincon:2015rma} for details and values of \mbox{the parameters}).

In the Lagrangian~(\ref{eq:lagPNJL}) the bare quark masses $m_{0i}$ are parameters. If these are taken to be zero, then the exact chiral symmetry will be spontaneously broken in vacuum by the quark condensate, and a set of (composed) Goldstone modes appears. When $m_{0i} \neq 0$ chiral symmetry is explicitly broken and after SSB, the masses of the pseudo-Goldstone bosons---in particular, pions and kaons---can be used to fix the values of these parameters (typically an $SU_I(2)$ isospin symmetry is still kept $m_{u0}=m_{d0}$, so that the three pions are degenerate in mass). 

Quark masses are dressed by interactions and modified by medium effects. At mean-field level, this is done through the quark condensates~\cite{Buballa:2003qv,Hansen:2006ee,Torres-Rincon:2015rma}
\be \label{eq:gap} m_i = m_{i0} - 4 G \langle \bar{\psi}_i \psi_i \rangle  + 2 H \langle \bar{\psi}_j \psi_j \rangle \ \langle \bar{\psi}_k \psi_k  \rangle \ , \quad j,k\neq i; j\neq k \ , \ee
where the quark condensate is given by
\be \langle  \bar{\psi}_i \psi_i \rangle = N_c \textrm{tr}_\gamma \sumint_q \frac{1}{\slashed{q}-m_i}\ , \ee 
where tr$_\gamma$ denotes the trace in Dirac space, and $\sumint_q$ is defined in Equation~(\ref{eq:sumint}).

Clearly the pseudo-Goldstone modes cannot be described by the degrees of freedom of the theory. However, mesonic candidates can be generated dynamically after solving the Bethe--Salpeter equation for the $\bar{q}q$ scattering. Consider the scattering problem for a quark-antiquark pair of flavors $i+\bar{j}$ to give $m+\bar{n}$. The equation for the $T$ matrix in the random-phase approximation in the imaginary time formalism reads~\cite{Vogl:1991qt,Klevansky:1992qe,Torres-Rincon:2015rma},
\be T^{ab}_{i \bar{j},m \bar{n}} (i\nu_m,{\bf p}) = {\cal K}^{ab}_{i \bar{j},m \bar{n}} - \sumint_k
{\cal K}^{ac}_{i \bar{j}, p \bar{q}} \  S_{p} \left( i\omega_n, {\bf k} \right) \ S_{ \bar{q}} \left( i\omega_n - i\nu_m, {\bf k}-{\bf p} \right)
\ T^{cb}_{p \bar{q},m \bar{n}} (i\nu_m,{\bf p}) \ , \label{eq:BSPNJL}
\ee
where $a,b$ denotes the flavor channel of the generated collective excitation. The kernel ${\cal K}$ contains color, flavor, and Dirac matrices,
\be {\cal K}^{ab}_{i\bar{j},m\bar{n}} = \Omega^a_{i \bar{j}} \ 2 K^{ab} \ \bar{\Omega}^b_{\bar{n}m} \ , \label{eq:Kcoupling} \ee
with
\be \Omega^a_{i \bar{j}}= \left( \unit_{\textrm{color}} \otimes \tau_{i \bar{j}}^a \otimes \Gamma \right)\ . \label{eq:omega} \ee

Given the Lagrangian (\ref{eq:lagPNJL}) the possible Dirac structures at the vertices are
\be \Gamma=\{1,i\gamma_5, \gamma^\mu, \gamma_5 \gamma^\mu \} \ . \ee

They can be used to generate scalar, pseudoscalar, vector, and axial-vector mesons, respectively.

The 't Hooft term of the Lagrangian breaks the flavor symmetry of the quartic coupling constant $G$. In the mean-field approximation, $H$ can be combined with $G$ to generate the effective couplings $K^{ab}$~\cite{Klevansky:1992qe,Rehberg:1995kh} 
\begin{align}
K^{00} &= G + \frac{H}{3} \left( \langle {\bar \psi_u} \psi_u \rangle+\langle  {\bar \psi_d} \psi_d\rangle  +\langle {\bar \psi_s} \psi_s \rangle \right) \ , \\
 K^{11} &= K^{22} = K^{33} = G - \frac{H}{2} \langle {\bar \psi_s} \psi_s \rangle \ , \\
K^{44} &= K^{55} = G - \frac{H}{2} \langle {\bar \psi_d} \psi_d\rangle  \ , \\ 
K^{66} &= K^{77}= G - \frac{H}{2} \langle {\bar \psi_u} \psi_u \rangle \ , \\
K^{88} &=  G - \frac{H}{6} \left(  2 \langle {\bar \psi_u} \psi_u \rangle +2 \langle  {\bar \psi_d} \psi_d \rangle- \langle {\bar \psi_s} \psi_s \rangle \right) \ , \\
K^{03} &=  K^{30} =  \frac{H}{2\sqrt{6}} \left(  \langle {\bar \psi_u} \psi_u \rangle -\langle {\bar \psi_d} \psi_d \rangle \right) \ , \\
K^{08} &=  K^{80} =  \frac{-H}{2\sqrt{6}} \left( \langle  {\bar \psi_u} \psi_u \rangle +\langle  {\bar \psi_d} \psi_d \rangle- 2 \langle  {\bar \psi_s} \psi_s \rangle \right) \ , \\
K^{38} &=  K^{83} =  -\frac{H}{2\sqrt{3}} \left( \langle {\bar \psi_u} \psi_u \rangle -\langle  {\bar \psi_d} \psi_d  \rangle \right) \ . 
\end{align}

There is a nondiagonal coupling in the (0-3-8) flavor channels, which describes the \mbox{$\pi^0-\eta^0-\eta^8$} mixing~\cite{Klevansky:1992qe}, which needs to be solved in the coupled-channel basis. Even in the isospin limit ($m_u=m_d$) a residual coupling
in the (0-8) channels remains, which eventually accounts for the mass difference between the $\eta$ and $\eta'$ mesons. In the absence of the 't Hooft term, these states would be degenerate.

Factorizing the function $t^{ab} $ as in Equation~(\ref{eq:Kcoupling}),
\be T_{i \bar{j},m \bar{n}}^{ab}  \equiv \Omega^a_{i \bar{j}} \ t^{ab}  \bar{\Omega}^b_{\bar{n}m} \ , \ee
the solution of Equation~(\ref{eq:BSPNJL}) reads
\be \label{eq:meson} t^{ab}  = \left[ \frac{2K}{1- 2 K \Pi} \right]^{ab} \ , \ee
where the polarization function $\Pi^{ab} (p^2)$ is calculated at finite temperature as
\be \label{eq:polmeson} \Pi^{ab} (i\nu_m,{\bf p}) = - \sumint_k \textrm{ tr}_{\gamma}
 \left[\bar{\Omega}^a_{\bar{j}i} \ S_i \left(i\omega_n , {\bf k} \right) \ \Omega^b_{i\bar{j}} \ S_{\bar j} \left( i\omega_n-i\nu_m ,{\bf k} - {\bf p} \right)  \right] \ , 
 \ee
where $S_i(i\omega_n,{\bf k})$ is the dressed quark propagator with flavor $i$ in the imaginary time formalism. The reduction of the polarization function in terms of simple integrals can be found in Reference~\cite{Rehberg:1995kh,Rehberg:1995nr,Torres-Rincon:2015rma}.

After analytic continuation to real energies, the poles of $t^{ab} (p_0+i\epsilon,{\bf p})$ represent dynamically generated mesonic states. Performing a Taylor expansion of  $t^{-1,ab} (p_0,0)$ around the pole $p_0=m_M$ one finds~\cite{Torres-Rincon:2015rma},

\be t^{ab} (p) \simeq  \frac{-g^2_{M\rightarrow \bar{q}q}}{p^2-m_M^2} \ , \ee
with the effective coupling
\be g^2_{M\rightarrow \bar{q}q} \equiv \frac{2m_M}{ \left. \frac{\pa \Pi^{ab} (p^2)}{\pa p} \right|_{p^2=m_M^2} } \ . \ee

Therefore, (the amputated) $t^{ab} (p^2)$ can be identified with the meson propagator in the appropriate spin-flavor channel, and the pole mass of the states can be computed via
\be \label{eq:mesonmass} 1- 2 K^{ab} \Pi^{ab} (p_0=m_M,{\bf p}=0) = 0 \ . \ee

If the emerging state has a $m_M$ larger than the sum of the masses of the two constituent quarks, then it is possible for this meson to decay into a $\overline{q}q$ pair---as no genuine confinement mechanism is imposed in the model. In this case Equation~(\ref{eq:mesonmass}) gives a complex solution. 
The real and imaginary parts of the solution $m_M$ are identified with the mass and (half) the decay width of the mesonic state, which become functions of temperature. I should mention that the complications of searching for poles in the different Riemann sheets is avoided after using an approximation which neglects the imaginary part of $p_0$ in some pieces of the polarization function~(\ref{eq:polmeson}). To the best of my knowledge this approximation was first used in~\cite{Zhuang:1994dw} and has been commonly used since then. Such an approximation should be acceptable for poles not far away from the real energy axis, while for states with a large decay width the analytic continuation to the unphysical Riemann sheet should be applied instead.

\begin{figure}[H]
\vspace{2pt}
\begin{center}
\includegraphics[width=0.36\textwidth]{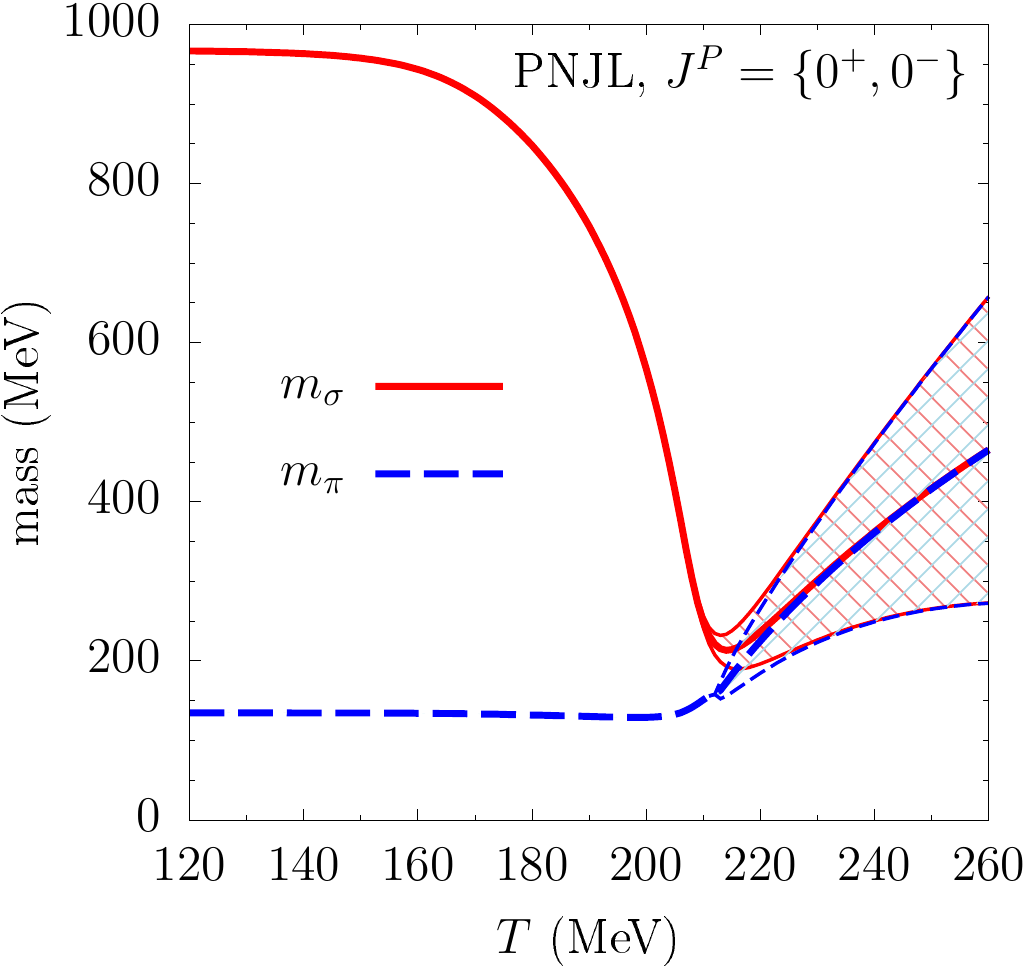}
\includegraphics[width=0.36\textwidth]{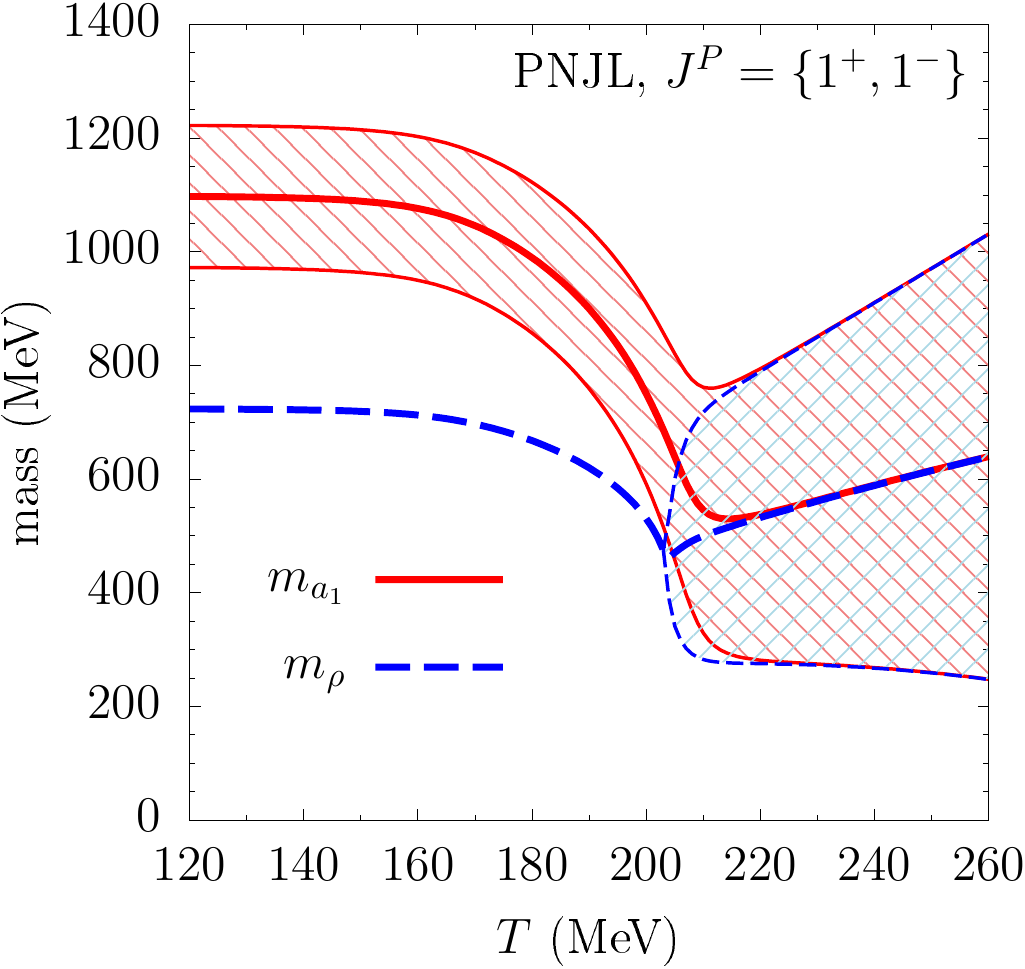}
\caption{(\textbf{Left panel}): $\pi$ and $\sigma$ masses as a function of temperature below and above the chiral phase transition of the PNJL model, using physical quark masses.  (\textbf{Right panel}): Same results for the vector-axial vector channel, the $\rho$ and $a_1$ mesons. The bands are a reflection of the thermal decay width of the different states, above the so-called Mott temperature, where the decay channel to a $\overline{q} q$ pair opens.\label{fig:PNJL1}}
\end{center}
\end{figure}

I present the results of the chiral partners masses in Figure~\ref{fig:PNJL1}. They are found by solving Equation~(\ref{eq:mesonmass}) in parity opposed channels, i.e., $J^P=0^+$ and $J^P=0^-$ but also $J^P=1^+$ and $J^P=1^-$. This is done by choosing the appropriate Dirac structure in the vertex \mbox{matrix~(\ref{eq:omega}).}
I used the PNJL model with the same parameters used in Reference~\cite{Torres-Rincon:2015rma}. The only difference with respect to that work is that the finite ultraviolet (UV) cutoff of the thermal integrals is removed as unnecessary. This treatment was in fact used later in Reference~\cite{Torres-Rincon:2017zbr}, as otherwise one cannot reproduce the Stefan--Boltzmann limit of the thermodynamic quantities for large temperatures~\cite{Zhuang:1994dw}. In the present context, the only effect is a reduction of the chiral transition temperature as can be seen by comparing the present results of Figure~\ref{fig:PNJL1} and the outcome of Reference~\cite{Torres-Rincon:2015rma}.

The left panel of Figure~\ref{fig:PNJL1} shows the scalar-pseudoscalar channel, i.e., the $\pi-\sigma$ case, with nonzero quark masses ($m_{u0}=m_{d0}=5.5$ MeV). The resulting thermal masses resemble the situation of the L$\sigma$M with $\epsilon\neq0$ shown in the right panel of Figure~\ref{fig:LSM}. However, there are some differences. First, the location of the chiral transition $T_c$ is very dependent on the model, the approximation used and the parameter set employed. Second, for the $SU_f(3)$ PNJL model, with the parameters used in Reference~\cite{Torres-Rincon:2015rma}, the vacuum mass of the generated state in the scalar-isoscalar channel is $m_\sigma=967$ MeV. Therefore, this state would be most likely to be identified with the $f_0 (980)$ rather than with the $f_0(500)$. The rather large vacuum quark masses (ca. 480 MeV for the light quark) can partially explain this result. No hints of a generated state close to the $f_0(500)$ was found in Reference~\cite{Torres-Rincon:2015rma}, pointing to a disfavored $\bar{q}q$ structure for this state. It would be interesting to re-analyze this scalar channel after coupling the $\pi-\pi$ scattering in the PNJL model, to see whether the $f_0(500)$ emerges once the pion-pion correlations are incorporated. 

 A thermal decay width for the two states is found at higher temperatures. This is indicated as a band around the respective thermal masses. Notice that the decay into a $q{\bar q}$ pair is a feature of the (P)NJL model but not physical due to the lack of real confinement in the model. The temperature at which the decay width becomes nonzero is called Mott temperature, and it is precisely the point in which the sum of the quark masses becomes equal to the real part of the pole position of the generated meson. It is interesting to observe that both the masses and the decay widths become degenerate around $T\simeq 220$ MeV, pointing to a degeneracy of the full spectral functions. The estimation of the chiral transition temperature, defined as the inflection point of the quark condensate, is $T_c \simeq 208$ MeV. Therefore, the effective restoration of chiral symmetry happens also above $T_c$, similarly to the L$\sigma$M shown in Section~\ref{sec:lsm}.  

 Because different Dirac structures have been allowed in the interaction terms of Lagrangian~(\ref{eq:lagPNJL}) one can easily compute meson thermal masses in different channels. Moving to the $J=1$ channel, degeneracy between vector and axial-vector states can be analyzed. In the right panel of Figure~\ref{fig:PNJL1} I present the results for the $\rho$ meson and its chiral partner the axial-vector $a_1$. Considering that a single coupling $G_V$ is used, the obtained vacuum masses of 723 MeV and 1098 MeV, respectively, are fair approximations to the experimental pole masses of 775 MeV and 1230 MeV~\cite{Zyla:2020zbs}, respectively. Notice that because the pion-pion channel is not coupled in the Bethe--Salpeter equation, the $\rho$ meson does not present the physical decay width to $\rho \rightarrow \pi+\pi$ in vacuum (see~\cite{He:1997gn} for the study of this particular process in the context of the NJL model). On the other hand, the $a_1$ mass at $T=0$ is above the two-quark mass threshold so that it presents a decay width at $T=0$. Again, above $T_c$ both masses and (large) decay widths do become degenerate.

\begin{figure}[H]
\begin{center}
\includegraphics[width=0.36\textwidth]{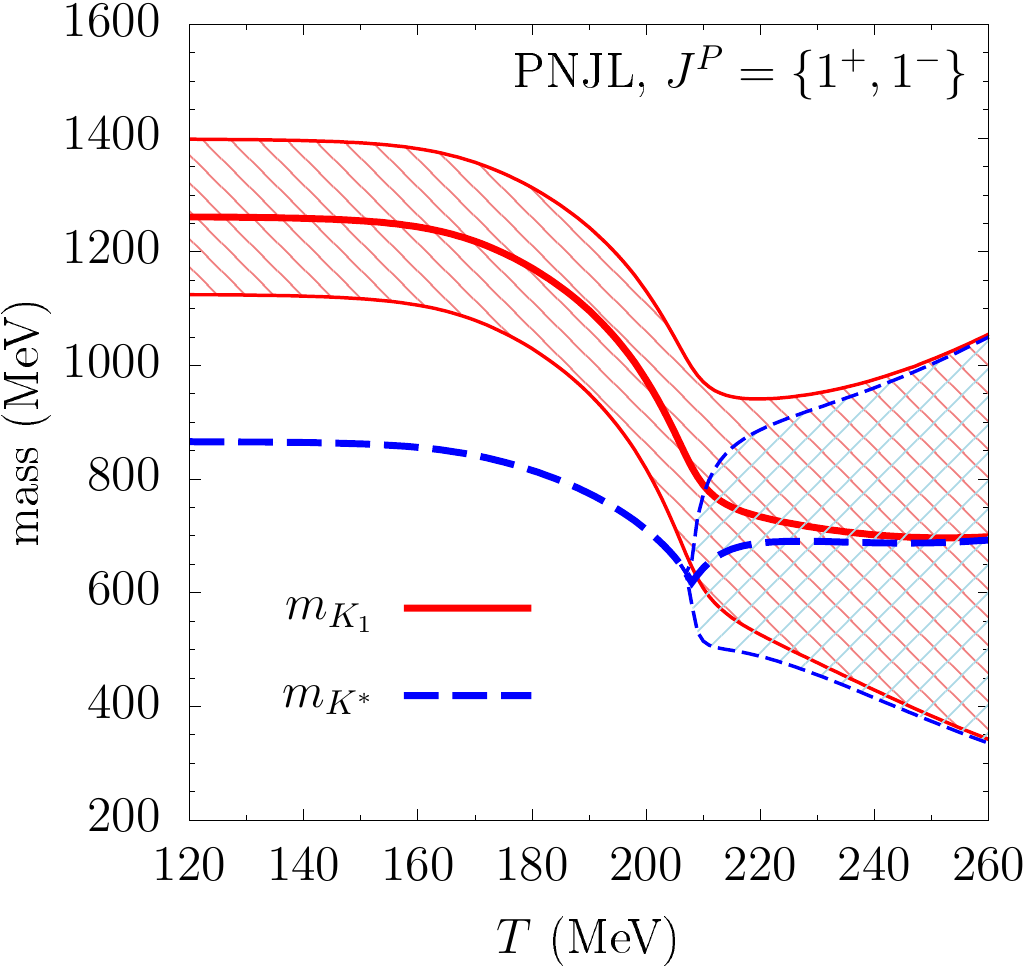}
\includegraphics[width=0.36\textwidth]{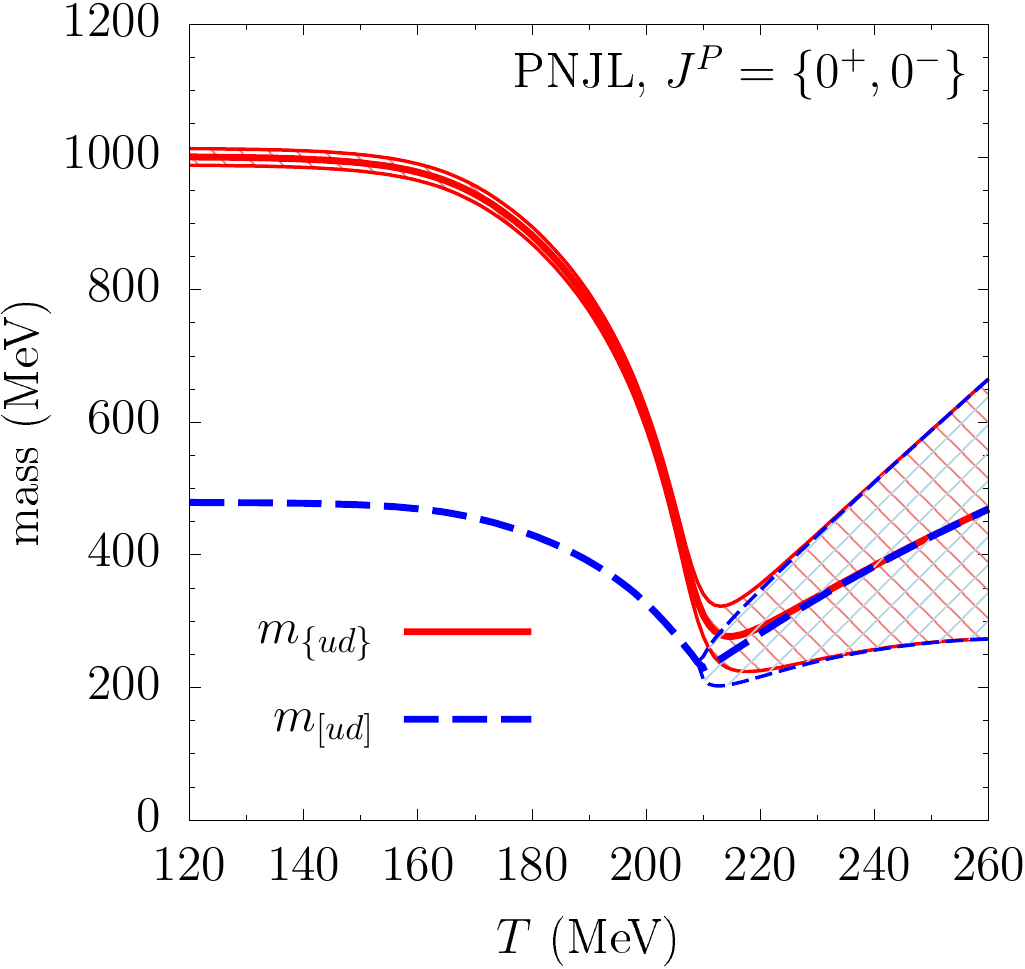}
\caption{(\textbf{Left panel}): Masses of $K^* (892)$ ($J^P=1^-$) and $K_1 (1270)$ ($J^P=1^+$) mesons as functions of temperature in the PNJL model below and above the chiral phase transition. (\textbf{Right panel}): Same results for the low-lying scalar ($[ud]$) and pseudoscalar ($\{ ud\}$) light diquarks.\label{fig:PNJL2}}
\end{center}
\end{figure}

\textls[-5]{In the strangeness sector a single additional parameter---the bare quark mass} $m_{s0}$---allows to access a series of new states~\cite{Torres-Rincon:2015rma}. I show an example of meson states containing net strangeness in the left panel of Figure~\ref{fig:PNJL2}, where I focus again on the vector channel (the pseudoscalar case was shown in Reference~\cite{Torres-Rincon:2015rma}). Again, the only difference with respect to that work is the suppression of the UV cutoff for the thermal integrations (the same value of the cutoff is kept for the vacuum contributions).

The vector state, with a vacuum mass of $m_{M}=866$ MeV, is associated to the $K^* (892)$ meson, while the axial-vector, and a mass of $m_M=1262$ MeV at $T=0$ is identified with the $K_1(1270)$~\cite{Zyla:2020zbs}. Notice that there are no free parameters to obtain these masses, as the vector coupling was already fixed in the nonstrange sector, and the current strange quark mass was fixed using the vacuum kaon mass~\cite{Torres-Rincon:2015rma}. The generated $K^*$ state shows no vacuum decay width at $T=0$, while the $K_1(1270)$ presents a decay probability to $\overline{q}q$. As for the $\rho$ meson case, it would be interesting to couple the $T$-matrix equation for the $K^*$ state to the $\pi-K$ scattering channel. However, within the (P)NJL model, the description of meson-meson scattering becomes rather involved~\cite{He:1997gn}.

To conclude this section I point out that the observed mass degeneracy at high temperature is not limited to $\overline{q}q$ states. In Reference~\cite{Torres-Rincon:2015rma} a similar Bethe--Salpeter equation was solved for the $qq$ scattering to generate diquark states. Those were used to model baryon states, upon the correlation with an additional quark. The diquark sector contains its free parameters in the form of unknown coupling constants. Even when these are in principle related to the ${\bar q}q$ couplings via Fierz transformation~\cite{Buballa:2003qv}, they are typically chosen as independent parameters to allow  for more flexibility. With two extra couplings (for the scalar and the axial-vector channels) it is possible to eventually explain the masses for all members of the baryon octet and decuplet representations of the $SU_f(3)$ group~\cite{Torres-Rincon:2015rma}. In the right panel of Figure~\ref{fig:PNJL2} I show the masses of the light diquark states in the scalar and pseudoscalar channels. The scalar diquark $[ud]$ is the lightest one and appears tightly bound with a mass less than 500 MeV in vacuum. The pseudoscalar diquark $\{ ud \}$ is very massive at $T=0$, with a small decay width. These states have opposite parity and become a degenerated chiral doublet above the chiral restoration point, as can be observed at high temperatures.

\section{Mixed Chiral Companions: Chiral Perturbation Theory~\label{sec:chpt}}

A third scenario of chiral degeneracy combines the previous two: one of the chiral partners is an explicit degree of freedom of the effective Lagrangian, while the other is a composed state emerging from the two-body dynamics. In this scenario the analysis of the mass evolution with temperature has to be performed between a thermal quasiparticle, whose vacuum mass is modified by thermal fluctuations, and a collective excitation, appearing as a dynamically-generated state.

Once more the $\pi$-$\sigma$ system can be studied under this category. The effective approach I will consider is the chiral perturbation theory (ChPT)~\cite{Gasser:1983yg,Gasser:1984gg,Leutwyler:1993iq}. This effective theory is considered the true effective description of QCD at low energies, as its Lagrangian is determined using all terms allowed by the symmetries of QCD. In addition, it incorporates the constraints from the chiral symmetry breaking pattern in a model independent way and exploits a well-defined power counting scheme~\cite{Weinberg:1978kz,Gasser:1983yg,Gasser:1984gg,Gasser:1986vb}. I refer to the excellent reviews~\cite{Leutwyler:1993iq,Meissner:1993ah,Ecker:1994gg,Pich:1995bw,Scherer:2002tk,Yndurain:2002ud,Scherer:2005ri} for details.

The degrees of freedom of ChPT are the lightest mesons (the pions for $N_f=2$, and the pseudoscalar meson octet for $N_f=3$). It is built following the pattern of the spontaneous chiral symmetry breaking $SU_L(N_f) \times SU_R(N_f) \rightarrow SU_V(N_f)$, where the degrees of freedom are related to the elements of the quotient group $SU_L(N_f) \times SU_R(N_f)/SU_V(N_f)$ (whose dimension is $N_f^2-1$). The lightest mesons are the Goldstone bosons appearing after the SSB. As in the previous examples, a finite quark mass can be introduced as an explicit chiral symmetry-breaking term, which---as being perturbatively small---is incorporated in the power counting of the theory~\cite{Gasser:1983yg,Gasser:1984gg}. 

The effective Lagrangian is expanded in even powers of derivatives over the Goldstone fields~\cite{Gasser:1983yg}
\be \label{eq:chiralexpansion} \mathcal{L}_{\textrm{ChPT}} = \mathcal{L}_2 + \mathcal{L}_4 + \mathcal{L}_6+ \cdots \ . \ee
where the lowest order Lagrangian coincides with the non-linear sigma model~\cite{Gasser:1983yg}
\be \mathcal{L}_2 = \frac{F_0^2}{4} \textrm{Tr } [ \pa_{\mu} U \pa^{\mu} U^{\dag} ] +\frac{F_0^2}{4} \textrm{Tr } [ \chi U^{\dag} + U \chi^{\dag} ] \ , \ee

For $N_f=2$, the exponential representation of fields $U(x)$ reads
\be U (x)= \exp \left( i \frac{\sqrt{2} \phi (x)}{ F_0} \right) \ , \label{eq:Umatrix} \ee
where 
\be \phi(x)= \lambda^a \phi_a (x) = \left(
\begin{array}{cc}
 \frac{1}{\sqrt{2}}\pi^0 & \pi^+  \\
 \pi^- & -\frac{1}{\sqrt{2}} \pi^0  \\
\end{array}
 \right) \ , \label{eq:phimatrix} \ee
and $\lambda^a$ ($a=1,2,3$) are the Pauli matrices.

The mass term for the pions contains the explicit symmetry-breaking term
\be \chi = 2B_0 \left(
\begin{array}{cc}
 m & 0 \\
0 & m
\end{array}
\right) = \left(
\begin{array}{cc}
 M_0^2 & 0 \\
0 & M_0^2
\end{array}
\right) \ , \ee
with $m=m_u \simeq m_d$ the light quark mass (in the isospin limit) and $M_0$ is the pion mass at lowest order. Finally, $F_0$ is a parameter which equals the pion decay constant at lowest order in the chiral limit~\cite{Gasser:1983yg}.

The next-order Lagrangian reads~\cite{Gasser:1983yg,Dobado:1997jx}
\begin{align}
\mathcal{L}_4 & = \frac{l_1}{4} \left\{ \textrm{Tr } [\pa^{\mu} U \pa_{\mu} U^{\dag}] \right\}^2 +\frac{l_2}{4} \textrm{Tr } [\pa_{\mu} U \pa_{\nu} U^{\dag}] \textrm{Tr } [\pa^{\mu} U \pa^{\nu} U^{\dag}] \nn \\
 & + \frac{l_3}{16} \left\{ \textrm{Tr } [\chi^{\dag} U + \chi U^{\dag}] \right\}^2 + \frac{l_4}{4} \textrm{Tr } [\pa_{\mu} U \pa^{\mu} \chi^{\dag} + \pa_{\mu} \chi \pa^{\mu} U^{\dag}] - \frac{l_7}{16} \left\{ \textrm{Tr } [ \chi U^{\dag}- U \chi^{\dag}] \right\}^2 \nn \\
 & + \frac{h_1+h_3}{4} \textrm{Tr } [\chi \chi^{\dag}] + \frac{h_1-h_3}{16} \left\{ (\textrm{Tr } [\chi U^{\dag}+ U \chi^{\dag}] )^2 +(\textrm{Tr } [\chi U^{\dag} - U \chi^{\dag}])^2 \right. \nn \\
 & \left. - 2 \textrm{Tr } [\chi U^{\dag} \chi U^{\dag} + U \chi^{\dag} U \chi^{\dag}] \right\} \ ,
\end{align}
where the constants $l_i$ and $h_i$ are called the ``low-energy constants'' and they are not known a priori as symmetry arguments alone do not fix them. They must be obtained from experiments or from lattice-QCD calculations.

Upon expansion of the unitary matrix of fields $U(x)$ one can describe interactions between pions at any order, according to the power-counting scheme. The partial scattering amplitudes $T_{IJ}$ at definite isospin $I$ and spin $J$ given from the ChPT Lagrangian are expressible as even powers of the pion momentum or in powers of the Mandelstam variable $s$ as,
\be T_{IJ}(s)=T_{IJ}^{(0)} (s) + T_{IJ}^{(1)} (s) + \cdots \ , \label{eq:TChPT} \ee
where $T_{IJ}^{(i)} (s)$ is $\mathcal{O} (s^{i})$. The partial amplitude $T_{IJ} (s)$ must fulfill the unitarity condition of the scattering ${\cal S}-$matrix. For the partial amplitude this condition reads (for the single channel case and above the two-meson energy threshold):
\be \label{eq:unitar} \textrm{Im } T_{IJ} (s)= \rho_{ab}(s) \ |T_{IJ} (s)|^2 \ , \ee
where $\rho_{ab}=\sqrt{[1-(m_a+m_b)^2/s][1-(m_a-m_b)^2/s]}$ is the two-meson phase space. However, the perturbative amplitudes obtained from ChPT only satisfy this relation in a perturbative way~\cite{Dobado:1992ha},
\begin{align} 
\textrm{Im } T_{IJ}^{(0)} & = 0 \ , \label{eq:imt0} \\ 
\textrm{Im } (T_{IJ}^{(0)} + T_{IJ}^{(1)}) & = \rho_{ab} \ |T_{IJ}^{(0)}|^2 \ , \label{eq:imt1} \\
\textrm{Im } (T_{IJ}^{(0)} + T_{IJ}^{(1)} + T_{IJ}^{(2)} ) & \simeq \rho_{ab} \ |T_{IJ}^{(0)} + T_{IJ}^{(1)}|^2  \ . \label{eq:imt2} 
\end{align}

The violation of the exact unitarity constraint~(\ref{eq:unitar}) produces an unnatural increase of the involved cross sections even at moderate energies (violation of unitarity bound~\cite{Yndurain:2002ud}). In addition, the polynomial expansion of the partial amplitudes (\ref{eq:TChPT}) makes it impossible to describe resonant interactions as they are unable to generate poles at any finite order in the expansion.
Some unitarization methods have been developed in order to cure this problem. I will first describe two of them, which I will exploit later in this work.

The inverse amplitude method (IAM)~\cite{Dobado:1989qm,Dobado:1992ha,Dobado:1996ps} is a way to construct a scattering amplitude that respects exact unitarity and is able to reproduce the presence of resonances as poles of the partial amplitudes. Omitting the details of the derivation (for which the reader can consult the mentioned references) the final amplitudes are approximated as (suppressing the isospin-spin indices to ease the notation)
\be \label{eq:iamamplit} T^{ \textrm{IAM}} (s) \simeq \frac{T^{(0),2} (s)}{T^{(0)} (s)-T^{(1)} (s)} \ , \ee
where the right-hand side is written in terms of known perturbative amplitudes~(\ref{eq:TChPT}), computed from ChPT at leading order (LO) $T^{(0)} (s)$ and at next-to-leading order (NLO) $T^{(1)} (s)$. The new scattering amplitude $T^{ \textrm{IAM}} (s) $ satisfies exact unitarity~(\ref{eq:unitar}), as
\be  \textrm{Im } T^{\textrm{IAM}} (s)= T^{(0),2} (s) \textrm{Im } \left[ \frac{1}{T^{(0)} (s)-T^{(1)}(s)} \right]  = \rho_{ab}(s) |T^{\textrm{IAM}} (s)|^2 \ , \label{eq:TIAM}\ee
where Equations~(\ref{eq:imt0}) and (\ref{eq:imt0}) were used.
Equation~(\ref{eq:TIAM}) being a rational function of $s$, it has the potential presence of poles which give access to resonant states in the relevant scattering channels in the appropriate Riemann sheet of the complex energy plane. One should mention that the basic IAM formula~(\ref{eq:TIAM}) has been extended in later works to accommodate several physical effects, thus improving its accuracy, e.g., by extending to coupled-channels with $N_f=3$~\cite{GomezNicola:2001as} or incorporating the presence of Adler zeros of the amplitudes~\cite{GomezNicola:2007qj}.

Finally I would like to point out the formal similarity between the  expressions in \mbox{Equations~(\ref{eq:iamamplit}) and~(\ref{eq:meson})}, with the identification $2K \leftrightarrow T^{(0)}(s)$ and $\Pi \leftrightarrow T^{(1)}/ T^{(0)}$. While the context and particular details of both expressions are rather different, in a profound sense both represent realizations of imposing the exact unitarity constraint of the ${\cal S}$ matrix.

The IAM has also been combined with other methods to improve the efficiency of the unitarization procedure. For example, in the $SU_f(3)$ ChPT unitarized amplitudes were obtained in~\cite{Oller:1997ng,Oller:1998hw} by combining the so-called ``on-shell factorization method''~\cite{Oller:1997ti} together with the IAM in a fully coupled-channel analysis. The unitarized amplitude $T^{\textrm{UChPT}} (s)$ has the form~\cite{Oller:1997ng,Oller:1998hw} (again suppressing $IJ$ indices),
\be T^{\textrm{UChPT}} (s)=T^{(0)}  [T^{(0)}-T^{(1p)}-T^{(0)} G T^{(0)}]^{-1} T^{(0)} \ , \label{eq:TUChPT} \ee
where $T^{(1p)}$ denotes the (polynomial) amplitude computed from the tree-level part of the NLO ChPT Lagrangian and the two-meson propagator $G$ reads,
\be G(s) = i \int \frac{d^4 k}{(2\pi)^4} \frac{1}{k^2-m_a^2+i\epsilon} \frac{1}{(P-k)^2 - m_b^2+i\epsilon} \ . \ee

The two-meson propagator is the analogue of the quark-antiquark propagator which was introduced in Equation~(\ref{eq:polmeson}) for the (P)NJL model. In fact, the imaginary part of $G$ is related two the  two-particle phase space,
\be \textrm{Im } G(s)=\rho_{ab}(s) \ . \ee

Together with the property $\textrm{Im } T^{(1p)} = 0$---as the polynomial part of the NLO amplitude is real---it is possible to show that the amplitude~(\ref{eq:TUChPT}) also satisfies the unitarity constraint in Equation~(\ref{eq:unitar}),
\be  \textrm{Im } T^{\textrm{UChPT}}= T^{(0)}  \textrm{Im } \left[ \frac{1}{T^{(0)} -T^{(1p)} -T^{(0)} G T^{(0)}} \right] T^{(0)}  = \rho_{ab}|T^{\textrm{UChPT}}|^2 \ . 
\ee

This concludes the basic elements of ChPT and the description of the unitarization methods which provide an improvement in the description of the scattering over the perturbative amplitudes. Now I turn to the description of the chiral partner masses at \mbox{finite temperature.}

  In thermal ChPT the pion mass dependence on temperature was addressed as early as~\cite{Gasser:1986vb,Gerber:1988tt} followed by the works~\cite{Schenk:1993ru,Song:1993ipa,Song:1994de,Toublan:1997rr}. In particular, in Reference~\cite{Song:1993ipa,Song:1994de} both the pole and screening pion masses were obtained using an effective approach up to temperatures of $T=150-200$ MeV. The effective Lagrangian of those works also included vector and axial-vector mesons as massive Yang-Mills fields, which modify the pion propagator. The net effect is a decrease of the pole mass of $\Delta m_\pi \simeq -20$ MeV at $T=200$ MeV, while the screening mass remains fairly constant. The thermal-averaged pion thermal width was estimated around 170 MeV at the same temperature~\cite{Song:1994de}, which supposes a reduction with respect to the predictions of lowest order ChPT for which the pion width goes like $\simeq\frac{T}{12} \left( \frac{T}{f_\pi} \right)^4$~\cite{Shuryak:1987ye,Goity:1989gs}.

ChPT was employed in Reference~\cite{Schenk:1993ru} to obtain a reduction of 60\% of the vacuum mass at $T=200$ MeV, a substantial decrease of the pion mass. A similar trend in the pion mass is seen in Reference~\cite{Toublan:1997rr}. However, these temperatures are too high to  rely on ChPT alone, which has a more restricted validity range (the typical pion momenta start to invalidate the power counting of the EFT). To improve this aspect, References~\cite{Schenk:1991xe,Schenk:1993ru} include some phenomenological fits to the $\pi-\pi$ scattering amplitudes based on experimental data. This was done to gauge the validity of standard ChPT at those temperatures. 

However---as I have already anticipated---an alternative possibility is to implement unitarized amplitudes based on $SU_f(3)$ ChPT. These can be used into the expression of the thermal self-energy provided in~\cite{Schenk:1991xe,Schenk:1993ru}. This was done, e.g., in Reference~\cite{FernandezFraile:2009kt} using the amplitudes~(\ref{eq:iamamplit}), and more recently in Reference~\cite{Montana:2020vjg} using the amplitudes~(\ref{eq:TUChPT}) from~\cite{Oller:1998hw}. I will provide some details on these calculations.

To address the thermal correction of the pion propagator one can exploit the expression obtained in~\cite{Schenk:1991xe,Schenk:1993ru} for the pion self-energy. Such calculation is based in a one-loop correction in the dilute limit. The thermal correction to the pion retarded self-energy reads,
\be
\Sigma^R (p^0,{\bf p})=-\int \frac{d^3q}{(2\pi)^3 2\omega_q} f(\omega_q,T) \overline{T_{\pi\pi}} (s) \ , \ee
where $\omega_q=\sqrt{{\bf q}^2+m_\pi^2(T=0)}$ is the free vacuum dispersion relation, $f(\omega_q,T)$ is the Bose--Einstein distribution function,
\be f(\omega_q,T) = \frac{1}{e^{\omega_q/T}-1} \ , \label{eq:BE}\ee
and $\overline{T_{\pi\pi}} (s)$ is the forward (isospin averaged) scattering amplitude of the $\pi\pi \rightarrow \pi\pi$ process.

The real part of the self-energy is responsible for the thermal correction to the vacuum pion dispersion relation,
\be \omega ({\bf p}) \simeq \omega_p - \frac{1}{2\omega_p}\int \frac{d^3q}{(2\pi)^3 2\omega_q}  f(\omega_q,T) \textrm{ Re } \overline{T_{\pi\pi}} (s) \ . \label{eq:pionomega} \ee

In the same approximation the on-shell thermal pion width $\Gamma( {\bf p})$ (which corresponds to the damping rate $\gamma( {\bf p})$ defined in~\cite{Schenk:1991xe,Schenk:1993ru}) is related to the imaginary part of the \mbox{retarded self-energy},
\be \Gamma( {\bf p}) = \frac{1}{\omega_p}\int \frac{d^3q}{(2\pi)^3 2\omega_q}  f(\omega_q,T) \textrm{ Im } \overline{T_{\pi\pi}} (s) \ . \label{eq:piongamma} 
\ee

Both quantities $\omega({\bf p})$ and $\Gamma({\bf p})$ can be represented together in the spectral function of the thermal pion, assuming that the quasiparticle picture is valid for the temperatures considered ($\Gamma/2 \ll \omega$). In this case the spectral function would read,
\be S_\pi (p^0,{\bf p})=\frac{1}{2\pi \omega ({\bf p})} \frac{\Gamma ({\bf p})/2}{[p^0-\omega({\bf p})]^2+[\Gamma ({\bf p})/2]^2} \ . \label{eq:pionspectral} \ee

Notice that the use of a unitarized $\pi\pi$ scattering amplitude (\ref{eq:TUChPT},\ref{eq:iamamplit}) in Equations~(\ref{eq:pionomega}) and (\ref{eq:piongamma}) is to be understood as a perturbative approach. The amplitudes are computed in vacuum and then inserted in the one-loop thermal corrections (a two-loop correction was also considered in~\cite{Schenk:1993ru}, but it would require the implementation of the six-pion scattering amplitude). Therefore, the main dependence on temperature is then assumed to be on the Bose--Einstein function Equation~(\ref{eq:BE}). Medium-dependent interactions would require a more sophisticated approach.

\begin{figure}[H]
\begin{center}
\includegraphics[scale=0.6]{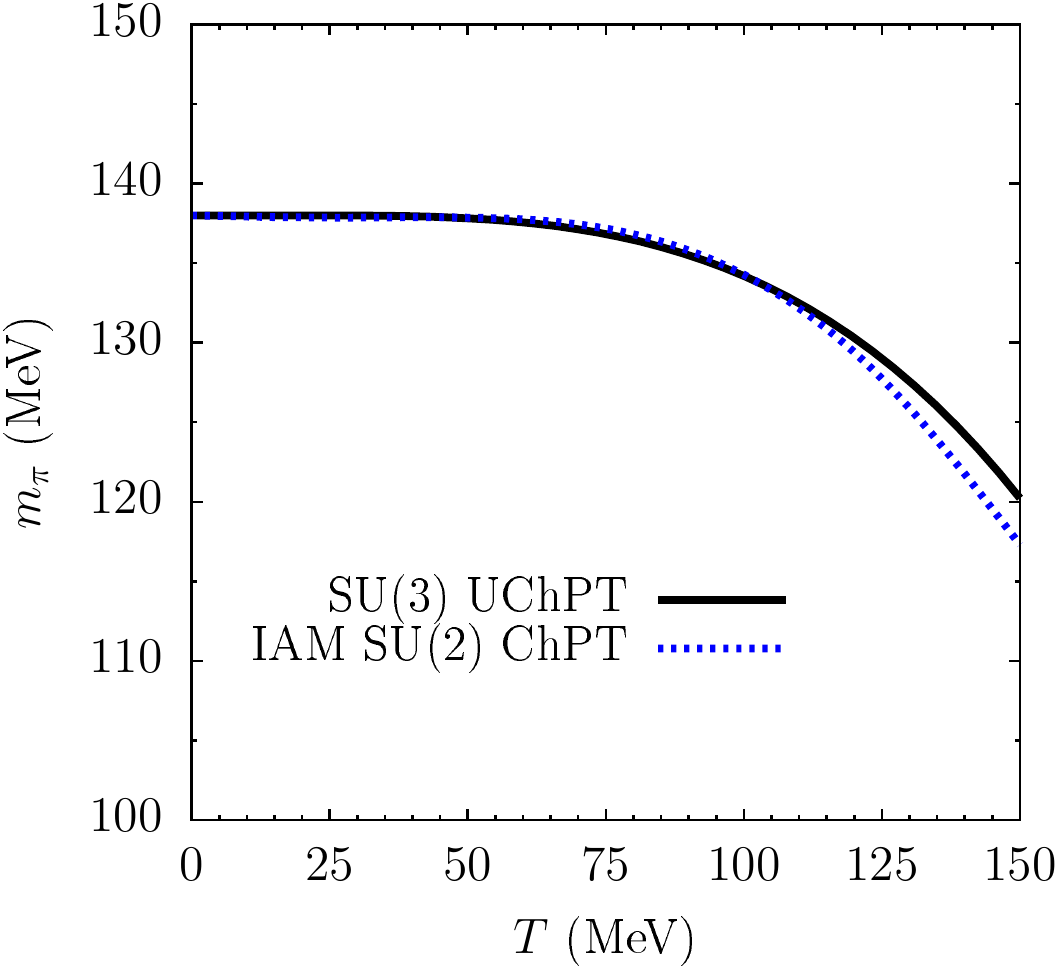}
\includegraphics[scale=0.6]{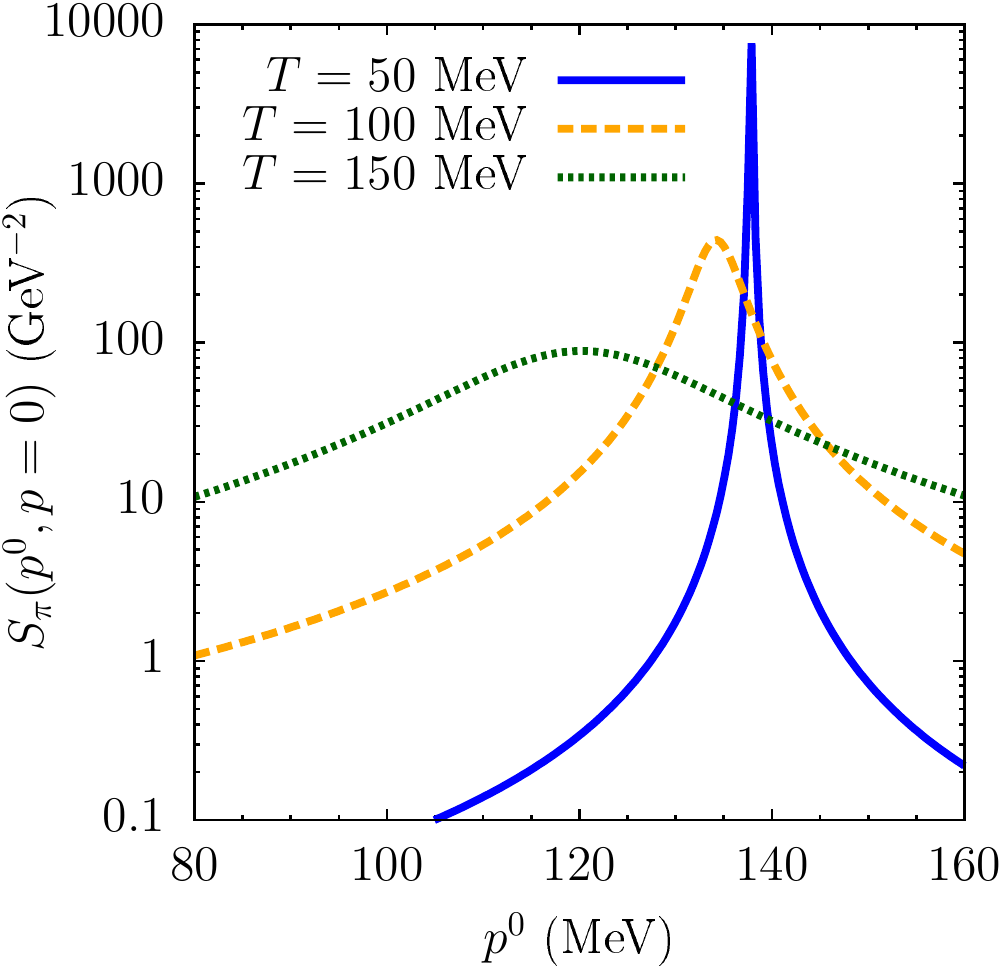}
\caption{(\textbf{Left panel}): Thermal pion mass obtained from the one-loop pion self-energy correction at finite temperature~\cite{Schenk:1993ru} using unitarized scattering amplitudes from ChPT. Solid line employs unitarized amplitudes from Reference~\cite{Oller:1998hw}, while dotted line uses the IAM results from Reference~\cite{FernandezFraile:2009kt}. (\textbf{Right~panel}): Spectral function~(\ref{eq:pionspectral}) of a pion at rest at different temperatures as a function of pion energy. Interactions are taken from Reference~\cite{Oller:1998hw}.~\label{fig:piChPT}}
\end{center}
\end{figure}

The thermal pion mass is computed as $m_\pi(T)=\omega({\bf p}=0)$ from Equation~(\ref{eq:pionomega}). It is plotted in the left panel of Figure~\ref{fig:piChPT} up to $T=150$ MeV. The result labeled ``SU(3) UChPT'' employs coupled-channel unitarized amplitudes from Reference~\cite{Oller:1998hw} (therefore, the mass modification contains subleading effects from higher mesonic states like the $K\bar{K}$ interaction). This case is taken from Reference~\cite{Montana:2020vjg}. With that interaction, at $T=150$ MeV the pion mass takes a value of $m_\pi \simeq 120$ MeV. Such reduction is not captured by the L$\sigma$M nor the (P)NJL model. Within ChPT this reduction was observed when implementing NLO amplitudes~\cite{Toublan:1997rr,FernandezFraile:2009kt}. In fact, the current algebra result does not show this trend~\cite{Schenk:1993ru} nor the LO ChPT amplitudes in the more recent Reference~\cite{Nicola:2014eda} where the pion mass always increases. This result for $m_\pi(T)$ is fully consistent with the one using the $SU_f(2)$ IAM in Reference~\cite{FernandezFraile:2009kt} at zero pion chemical potential, which is added to the same panel of \mbox{Figure~\ref{fig:piChPT}} under the label ``IAM SU(2) ChPT''. Apart from the intrinsic details of the different methods, the small differences at high temperatures could be ascribed to the absence of the $K\bar{K}$ channel in the second calculation.

In the right panel of Figure~\ref{fig:piChPT} I present the pion spectral function in the quasiparticle approximation~(\ref{eq:pionspectral}) using $\omega({\bf p})$ and $\Gamma({\bf p})$ in Equations~(\ref{eq:pionomega}) and (\ref{eq:piongamma}), respectively. The interactions are those described by the ``SU(3) UChPT'' amplitudes, cf. Equation~(\ref{eq:TUChPT}). The pion is taken at rest and the spectral function is plotted as a function of the energy for three different temperatures $T=50,100,150$ MeV. One can observe the reduction of the pion thermal mass with temperature in the shift of the quasiparticle peak as well as the increase of the thermal decay width from the broadening of the spectral function. At \mbox{$T=150$ MeV} the decay half-width is $\Gamma( {\bf p}=0)/2 \simeq 15$ MeV, which is still one order of magnitude smaller than $\omega ({\bf p}=0) \simeq 120$ MeV, and therefore the quasiparticle approximation should be reasonable, at least for low-momentum pions.

Unfortunately ChPT cannot be applied much beyond $T=150$ MeV, and therefore it is not possible to study the properties of pions above the chiral restoration temperature. According to the L$\sigma$M and the (P)NJL, only for $T>T_c$ the true degeneracy of states takes place. Nevertheless, I will therefore try to analyze any indication of chiral partner degeneracy below $T_c$. For this I need to describe the emergence of the $J^P=0^+$ chiral companion of the pion in this approach.

It should be mentioned that the history of the $\sigma$ state (or $f_0(500)$ according to the Particle Data Group~\cite{Zyla:2020zbs}) is a long-standing one (and sometimes polemic). For a comprehensive discussion I refer the reader to the detailed review~\cite{Pelaez:2015qba}. From the latest version of the Particle Data Group review~\cite{Zyla:2020zbs} it is a state with a mass between 400 and 550 MeV and a large decay width. More precisely, the value of the imaginary part of the pole position in the complex energy plane is estimated as (200--350) MeV (with a full width estimated using a Breit-Wigner ansatz twice this value $\Gamma=(400-700)$ MeV). Therefore, it is very broad state already at $T=0$. 

The $\sigma$ is not part of the ChPT degrees of freedom, but it could be generated via the correlation of two pions, to which the $\sigma$ can decay in an $s-$wave with a large probability~\cite{Meissner:1990kz}. In fact, it is possible to describe this state by unitarizing the scalar-isoscalar scattering channel of two pions from ChPT~\cite{Dobado:1989qm,Dobado:1992ha,Dobado:1996ps,Oller:1997ti}. In particular, this was achieved in the studies~\cite{Oller:1998hw} and~\cite{FernandezFraile:2009kt}, which I have already used to generate the results of Figure~\ref{fig:piChPT}.

\begin{figure}[H]
\begin{center}
\includegraphics[width=0.35\textwidth]{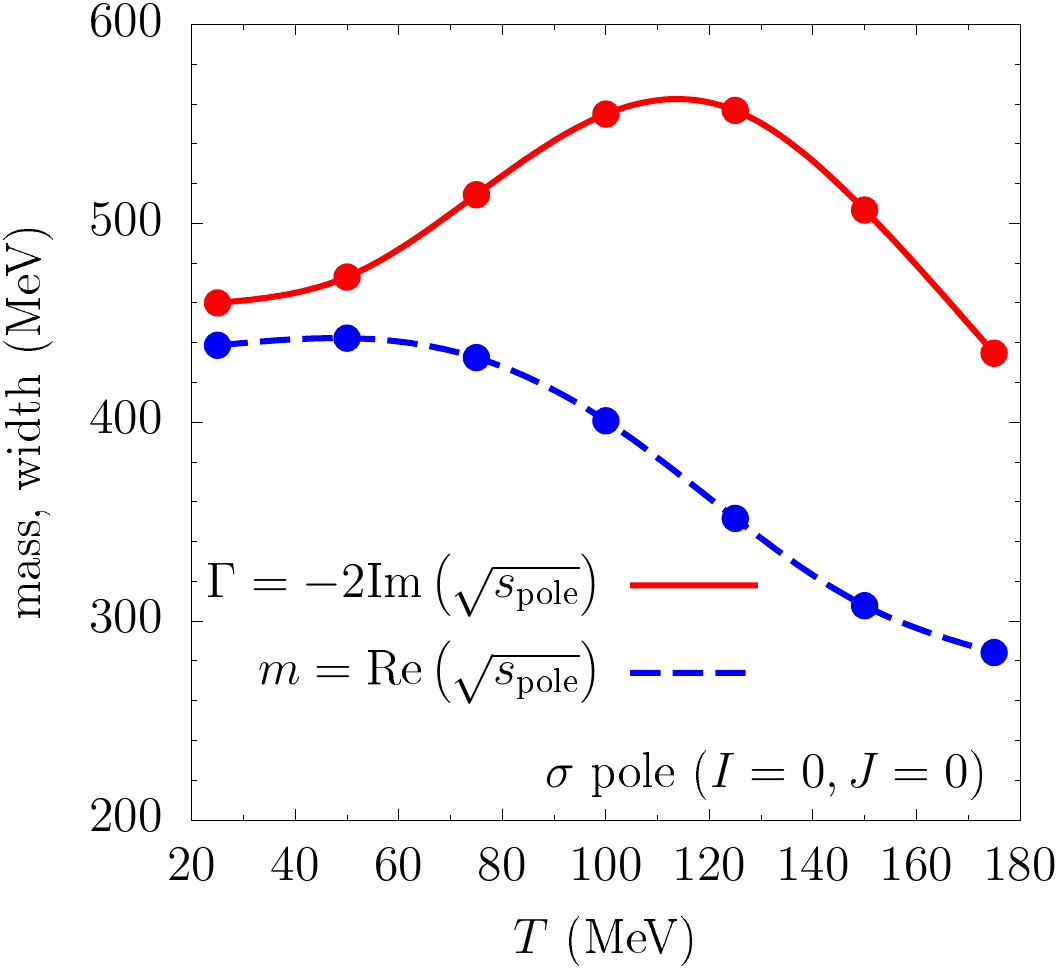}
\caption{Mass and decay width of the $\sigma$ resonance as obtained in the IAM method applied to $SU_f(3)$ ChPT at finite temperature~\cite{Dobado:2002xf,GomezNicola:2002an}. The mass is computed as the real part of the pole position in the unphysical Riemann sheet, while the decay width is defined as twice the absolute value of the imaginary part of the pole position.~\label{fig:sigmaChPT}}
\end{center}
\end{figure}

Unfortunately, studies on the generation of the $f_0(500)/\sigma$ at finite temperature in the context of the ChPT are scarce. A calculation of the thermal mass/decay width of this state can be found in References~\cite{Dobado:2002xf,GomezNicola:2002an}, where the coupled-channel version of the $SU_f(3)$ IAM is employed. There, the unitarized scattering amplitudes are extended to include medium effects, together with thermal modification of the kinematic phase-space. The vacuum value of the $\sigma$ pole mass was estimated as $\simeq$440 MeV and its decay width around 460 MeV. When the temperature increases, the pole mass decreases to 350 MeV at $T=125$ MeV and the width increases up to 560 MeV. The results of References~\cite{Dobado:2002xf,GomezNicola:2002an} are plotted in Figure~\ref{fig:sigmaChPT} for the mass and the decay width of the $\sigma$ resonance.

The behavior of the thermal $\sigma$ is also described in the more recent~\cite{Ferreres-Sole:2018djq}, which also implements the IAM to dynamically generate the $\sigma$ pole from the $\pi\pi$ scattering. In that work, the $\sigma$ mass has a somewhat stronger reduction reaching at $T=150$ MeV half of its vacuum mass. From the results of~\cite{Ferreres-Sole:2018djq}, and assuming a vacuum $\sigma$ mass of 440 MeV, together with a pion mass of 120 MeV at the same temperature (according to the results in Figure~\ref{fig:piChPT}), the mass difference of the chiral states is $m_\sigma-m_\pi \simeq 100$ MeV (for comparison, at $T=0$ this difference is $\simeq$300 MeV). Therefore, one can conclude that there exist some indications that the chiral mass degeneracy could be partially realized in ChPT as well. However, notice that the $\sigma$ decay width is likely to be rather large at $T=150$ MeV (similar to what is shown in Figure~\ref{fig:sigmaChPT}), while the pion only gets a couple of dozens of MeV of width at the same temperature, as I discussed before. This observation seems to preclude a full degeneracy of the spectral functions of the chiral partners at high temperature. Nonetheless, it would be necessary to carry out a self-consistent calculation in which the pion is sensitive to the $\sigma$ mass reduction in medium. This reduction might have a potential effect in the pion thermal decay width.      

To conclude this section I point out that in the chiral limit ($m_u=m_d=0$) and argument for the $\pi-\sigma$ mass degeneracy can be found in Reference~\cite{Oller:2000wa}, based on the reduction of the pion decay constant $f_\pi$ with temperature~\cite{Gasser:1986vb,Bochkarev:1995gi,Pisarski:1996zv}. The reduction of $f_\pi$ in a dense medium~\cite{Meissner:2001gz,Weise:2001sg} has also been exploited in the work~\cite{Yokokawa:2002pw} to compute density modifications of the $\pi\pi$ scattering amplitude of ChPT in the $\rho$ and $\sigma$ meson channels. I will insist more on the effect of $f_\pi$ at the end of the next section.

\section{Chiral Partner as a Double-Pole Structure: Open Charm Mesons~\label{sec:heavy}}

In this section I will consider a variation of the previous situation, where the positive-parity chiral companion is generated via the dynamics of the degrees of freedom and emerges as a new, resonant state. The difference lies in the fact that this state is not represented with a single pole of the scattering amplitude but with a two-pole structure. I will focus on the charm sector and describe the situation of $D$ mesons at finite temperature, as presented in References~\cite{Montana:2020lfi,Montana:2020vjg}.

In the last decade of heavy-flavor systems have received a lot of attention thanks to the development of effective theories describing charm and bottom. The large masses of these states compared to other scales in the system help to define a power counting to apply a suitable expansion of the effective Lagrangian. Some effective approaches in the heavy-flavor sector are the ``heavy-quark effective theory''~\cite{Isgur:1989vq,Eichten:1989zv,Georgi:1990um,Neubert:1993mb} and the ``non-relativistic QCD'' and its extensions~\cite{Caswell:1985ui,Bodwin:1994jh,Manohar:1997qy,Pineda:1997bj,Brambilla:1999xf}. In addition, open charm and bottom mesons were studied using a combination of heavy-quark spin-flavor symmetry and chiral symmetry~\cite{Burdman:1992gh,Wise:1992hn,Yan:1992gz} in the recent works~\cite{Kolomeitsev:2003ac,Lutz:2007sk,Guo:2008gp,Guo:2009ct,Geng:2010vw,Abreu:2011ic}.

 Let me briefly describe the spectroscopy situation of the open charm states in vacuum~\cite{Zyla:2020zbs}. Focusing on the low-lying states, the ground state $D$ with a (isospin average) mass of 1867 MeV is a pseudoscalar ($J^P=0^-$) as well as the strange counterpart $D_s$ (with a mass of 1968 MeV). The heavy-quark flavor partner of the $D$ meson is the $\bar{B}$ meson, and its heavy-quark spin partner is the vector $D^*$ ($m_{D^*}=2008$ MeV). At the level of the EFT at LO, the $D$ meson is fully decoupled from the $D^*$ and they are described by analogous interactions (except for the explicit breaking when using the physical masses)~\cite{Geng:2010vw,Abreu:2011ic}.

\begin{figure}[H]
\begin{center}
\includegraphics[scale=0.65]{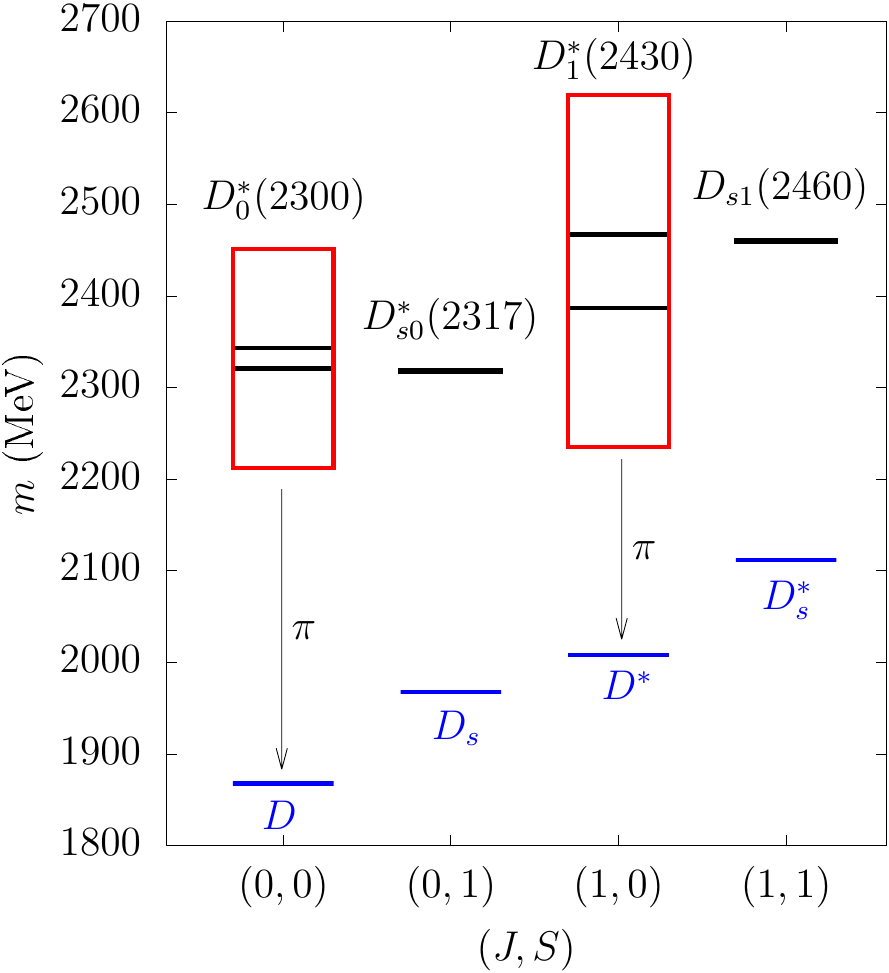}
\caption{Low-lying $D$-meson states in the $J=\{0,1\}$ and $S=\{0,1\}$ channels according to the Particle Data Group review~\cite{Zyla:2020zbs}. The height of the black boxes represents the uncertainty in the masses of the states, while the height of the red boxes is a measure of the vacuum decay width of the resonances.~\label{fig:spectrum}}
\end{center}
\end{figure}

The chiral partner candidate of the $D$ meson is a $J^P=0^+$ state with a mass around 2 GeV called $D_0^*(2300)$~\cite{Zyla:2020zbs}.  The $D_0^*(2300)$ decays into the ground state emitting a pion in $s-$wave~\cite{Zyla:2020zbs}. Therefore, there is a chance to describe this state from the $D$-$\pi$ two-body problem (similar to the $\sigma$ resonance in the $\pi\pi$ interaction). The analogue companion of the $D_s$ is the spin-0 $D_s^*(2317)$~\cite{Bardeen:2003kt}, which is a very narrow state (an upper limit of the decay width is 3.8 MeV~\cite{Zyla:2020zbs}). In the vector sector, the chiral partner of the $D^*$ is the broad $D^*_{1} (2430)$, and the companion of the $D_s^*$ is the narrow $D_{s1} (2460)$. A summary of the different states according to their total angular momentum ($J$) and their strangeness ($S$) content is given in Figure~\ref{fig:spectrum}.

Anticipating the situation for the $D_0^*(2300)$ resonance, what makes this state particularly interesting is its two-pole structure~\cite{Kolomeitsev:2003ac,Guo:2006fu,Guo:2009ct,Albaladejo:2016lbb} (see also the recent review~\cite{Meissner:2020khl}). The physical state quoted in~\cite{Zyla:2020zbs} is therefore a combination of two poles (sharing the same quantum numbers) located at different positions in the complex-energy plane. The quoted mass of the $D_0^*(2300)$ is $(2318\pm 30)$ MeV, sitting in between the respective real parts of the two poles~\cite{Albaladejo:2016lbb}. The decay width is large $\Gamma=261\pm40$ MeV and similar to (twice) the imaginary part of the two pole positions. 

I now detail how the states shown in Figure~\ref{fig:spectrum} are generated in vacuum and at finite temperature. In References~\cite{Montana:2020lfi,Montana:2020vjg} we exploited the effective approach of References~\cite{Lutz:2007sk,Guo:2008gp,Guo:2009ct,Geng:2010vw} and extended it to finite temperature using the imaginary time formalism~\cite{Kapusta:2006pm}. The effective Lagrangian is kept at leading-order (LO) in heavy-quark mass expansion but at next-to-leading (NLO) order in chiral expansion. The LO Lagrangian contains the Goldstone-boson sector (that is, the ChPT Lagrangian at LO) plus the sector involving heavy mesons. These are encoded as an antitriplet in $D=\begin{pmatrix} D^0 & D^+ & D^+_s \end{pmatrix}$, and their heavy-quark partners, the vector $D^*_\mu=\begin{pmatrix} D^{*0} & D^{*+} & D^{*+}_s \end{pmatrix}_\mu$. The octet of pseudo-Goldstone bosons follows the exponential representation $u(x)=\exp (i\Phi(x)/\sqrt{2} f_\pi)$, where 

\be \Phi(x) = \left(
\begin{array}{ccc}
\sqrt{\frac12} \pi^0 +\sqrt{\frac16} \eta & \pi^+ & K^+ \\
\pi^- & -\sqrt{\frac12} \pi^0+\sqrt{\frac16} \eta & K^0 \\
K^- & \bar{K^0} & -\sqrt{\frac23} \eta \\
\end{array}
\right)\ . \ee

Notice that the matrix $U$ defined in Equation~(\ref{eq:Umatrix}) is equivalent to $u^2$ introduced here, and the field matrix $\phi$ in Equation~(\ref{eq:phimatrix}) is the 2-flavor equivalent of the $\Phi$ shown here. The LO Lagrangian reads
\begin{align}
 \mathcal{L}_{\rm LO}&\ =\mathcal{L}^{\rm ChPT}_{\rm LO}+\langle\nabla^\mu D\nabla_\mu D^\dagger\rangle-m_D^2\langle DD^\dagger\rangle-\langle\nabla^\mu D^{*\nu}\nabla_\mu D^{*\dagger}_{\nu}\rangle+m_D^2\langle D^{*\nu}D^{*\dagger}_{\nu}\rangle  \nn \\
 &\ +ig\langle D^{*\mu}u_\mu D^\dagger-Du^\mu D^{*\dagger}_\mu\rangle+\frac{g}{2m_D}\langle D^*_\mu u_\alpha\nabla_\beta D^{*\dagger}_\nu-\nabla_\beta D^*_\mu u_\alpha D^{*\dagger}_\nu\rangle\epsilon^{\mu\nu\alpha\beta} \ ,
 \label{eq:lagrangianLO}
\end{align}
where $u_\mu=i(u^\dagger\partial_\mu u-u\partial_\mu u^\dagger)$, $\nabla_\mu D^{(*)}=\partial_\mu D^{(*)} -D^{(*)}\Gamma_\mu$, and the connection $\Gamma_\mu=\frac{1}{2}(u^\dagger\partial_\mu u+u\partial_\mu u^\dagger)$. The angular brackets imply a trace in flavor space.

At NLO the Lagrangian contains several low-energy constant $h_i,\tilde{h}_i$~\cite{Guo:2009ct,Geng:2010vw,Abreu:2011ic,Liu:2012zya,Tolos:2013kva},
\begin{align}
 \mathcal{L}_{\rm NLO}=&\ \mathcal{L}^{\rm ChPT}_{\rm NLO} -h_0\langle DD^\dagger\rangle\langle\chi_+\rangle+h_1\langle D\chi_+D^\dagger\rangle+h_2\langle DD^\dagger\rangle\langle u^\mu u_\mu\rangle \nn \\ 
 &\ +h_3\langle Du^\mu u_\mu D^\dagger\rangle+h_4\langle\nabla_\mu D\nabla_\nu D^\dagger\rangle\langle u^\mu u^\nu\rangle+h_5\langle\nabla_\mu D\{u^\mu,u^\nu\}\nabla_\nu D^\dagger \rangle \nn \\ 
 &\ +\tilde{h}_0\langle D^{*\mu}D^{*\dagger}_\mu\rangle\langle\chi_+\rangle-\tilde{h}_1\langle D^{*\mu}\chi_+D^{*\dagger}_\mu\rangle-\tilde{h}_2\langle D^{*\mu}D^{*\dagger}_\mu\rangle\langle u^\nu u_\nu\rangle \nn \\ 
 &\ -\tilde{h}_3\langle D^{*\mu}u^\nu u_\nu D^{*\dagger}_\mu\rangle-\tilde{h}_4\langle\nabla_\mu D^{*\alpha}\nabla_\nu D^{*\dagger}_\alpha\rangle\langle u^\mu u^\nu\rangle-\tilde{h}_5\langle\nabla_\mu D^{*\alpha}\{u^\mu,u^\nu\}\nabla_\nu D^{*\dagger}_\alpha\rangle \ , \label{eq:lagrangianNLO}
\end{align}
where ${\mathcal L}^{\rm ChPT}_{\rm NLO}$ represents the NLO ChPT Lagrangian involving only $\Phi$, and $\chi_+=u^\dagger\chi u^\dagger+u\chi u$ with the quark mass matrix $\chi={\rm diag}(m_\pi^2,m_\pi^2,2m_K^2-m_\pi^2)$. 

The tree-level amplitudes from the LO+NLO Lagrangian at lowest order in the inverse $D$-meson mass expansion are~\cite{Guo:2009ct,Geng:2010vw},

\begin{align}
 V^{ij}(s,t,u) &=  \frac{1}{f_\pi^2}\Big[\frac{C_{\rm LO}^{ij}}{4}(s-u)-4C_0^{ij}h_0+2C_1^{ij}h_1  \nn \\ 
 & -2C_{24}^{ij}\Big(2h_2(p_2\cdot p_4)+h_4\big((p_1\cdot p_2)(p_3\cdot p_4)+(p_1\cdot p_4)(p_2\cdot p_3)\big)\Big)  \nn \\
 &\ +2C_{35}^{ij}\Big(h_3(p_2\cdot p_4)+h_5\big((p_1\cdot p_2)(p_3\cdot p_4)+(p_1\cdot p_4)(p_2\cdot p_3)\big)\Big) \Big] \ , \label{eq:potential}
\end{align}
where $s=(p_1+p_2)^2,t=(p_1-p_3)^2,u=(p_1-p_4)^2$ are the Mandelstam variables. The indices $i,j$ represent initial and final scattering channels, and $C^{ij}$ are isospin coefficients depending on the scattering channels, which are described in detail in~\cite{Montana:2020lfi,Montana:2020vjg}. Both elastic and inelastic channels (like $D\pi \rightarrow D_s {\bar K}$) are incorporated in the framework.

Similar to the ChPT case, these amplitudes are only valid at low momentum, violating the exact unitarity constraint~(\ref{eq:unitar}) at moderate energies. However, the attractive nature of the interaction calls for a resummation in the form of a Bethe--Salpeter equation. Its solution provides a set of unitarized scattering amplitudes. The motivation for such an approach was explained in Section~\ref{sec:chpt} and also applies in the context of the heavy-flavor sector. In the so-called ``on-shell factorization method''~\cite{Oller:1997ti,Oset:1997it} the $T$-matrix equation reads,
\be \label{eq:BetheSalpeter} T^{ij} (s)= V^{ij} (s) + V^{il}(s) G^l(s) T^{lj} (s) \ , \ee
where a sum over all possible intermediate channels $l$ is implied. $V^{ij}(s)$ corresponds to the $s$-wave projection of the perturbative amplitude~(\ref{eq:potential}). In vacuum, the two-propagator function is computed as
\be\label{eq:loopVac}
  G^l(s)=i\int\frac{d^4q}{(2\pi)^4} \frac{1}{q^2-m_D^2+i\epsilon} \frac{1}{(p-q)^2-m_\Phi^2+i\epsilon} \ ,
\ee
with $p^\mu=(E,{\bf p})$. I make explicit that at $T=0$ the loop function is given as a function of the Mandelstam variable $s=p^2=(p_1+p_2)^2$. The solution of the $T$-matrix equation is
\be T_{ij}(s)= V_{ik}(s) [1-V(s)G(s)]^{-1}_{kj} \ , \label{eq:TmatrixD} \ee
which satisfies the exact unitarity relation of the scattering matrix~(\ref{eq:unitar}). Again I stress the formal similarity of this equation with Equation~(\ref{eq:meson}) where the role of the perturbative amplitude $V$ is equivalent to the NJL 4-quark coupling $2K$, and the two-meson propagator $G$ corresponds to the polarization function ($q{\bar q}$ propagator) $\Pi$.

\begin{figure}[H]
\begin{center}
\includegraphics[width=0.36\textwidth]{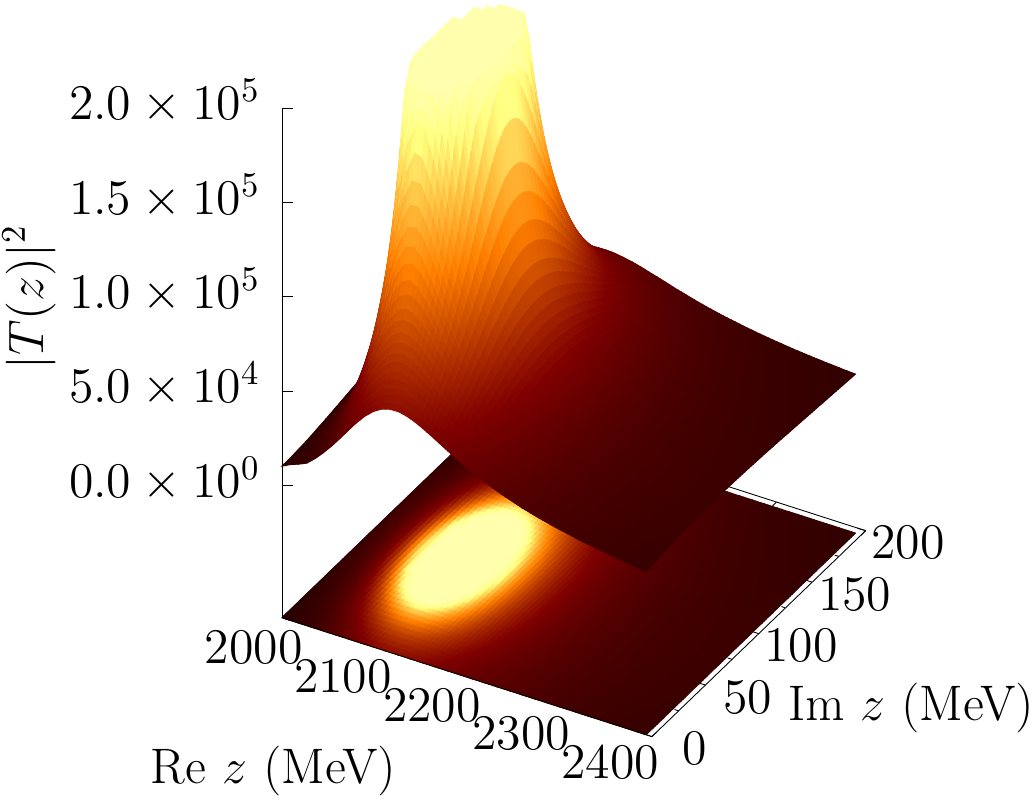}
\includegraphics[width=0.36\textwidth]{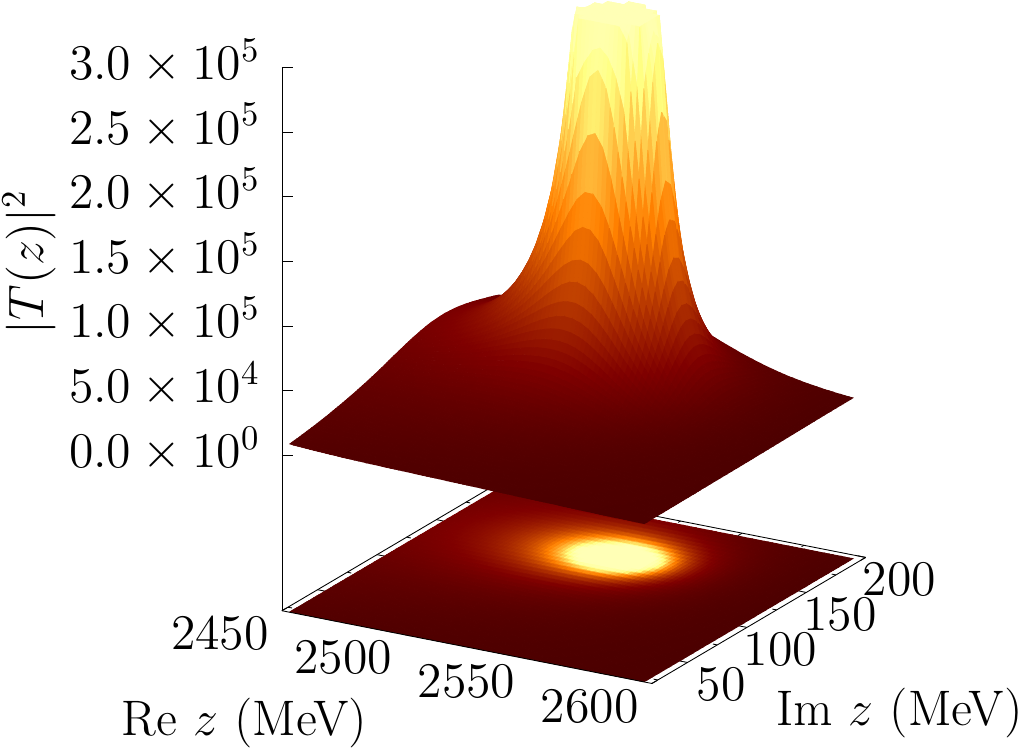}
\caption{Two poles of the $D_0^*(2300)$ at $T=0$ as they appear in the complex energy plane of the $D\pi \rightarrow D\pi$ channel in Reference~\cite{Montana:2020vjg}. (\textbf{Left panel}): $(-,+,+)$ Riemann sheet where the lower pole emerges. (\textbf{Right panel}):~$(-,-,+)$ Riemann sheet where the upper pole appears.~\label{fig:polesD0}}
\end{center}
\end{figure}

Equation~(\ref{eq:TmatrixD}) is solved in the different channels of angular momentum ($J$) and strangeness ($S$). For $(J,S)=(0,0)$ we found that the two poles of the $D_0^*(2300)$ state are located in two different Riemann sheets: the lower pole in the $(-,+,+)$ sheet in a position $\sqrt{s_{\textrm{pole}}} = (2082,86)$ MeV; whereas the higher one is located in the $(-,-,+)$ sheet at $\sqrt{s_{\textrm{pole}}}=(2526,147)$ MeV. These poles can be observed in Figure~\ref{fig:polesD0} where I plot the different Riemann sheets of the diagonal $D\pi \rightarrow D\pi$ channel. The lower pole appears in a Riemann sheet which is directly connected to the physical Riemann sheet below the real energy axis and therefore it is very prominent along the real axis (physical scattering amplitude). The upper pole appears in a RS which is not directly connected with the physical RS. In addition, it has a large imaginary part, so its effect will not be significant in the physical scattering amplitude (however, as shown in Reference~\cite{Montana:2020vjg}, it plays a relevant role in the $D_s \bar{K}$ channel, to which it couples more strongly).

In the $(J,S)=(0,1)$ channel a single pole below the $DK$ threshold is found in the unphysical Riemann sheet $(+,+)$ with a mass of 2252.5 MeV. This state is identified with the narrow state $D_{s0}^* (2317)$. While the mass is somewhat below the mass listed in~\cite{Zyla:2020zbs}, we found some sensitivity with the parameter set, as explained in Reference~\cite{Montana:2020vjg}. I should mention that according to Reference~\cite{Zyla:2020zbs} this state has a small decay width to $D\pi$, which is a process that does not conserve isospin. Including an isospin breaking term in the EFT, one can explain such decay~\cite{Cho:1994zu}.

At finite temperature one can perform a similar situation using the imaginary-time formalism~\cite{Kapusta:2006pm}. As opposed to the calculation presented in Reference~\cite{Schenk:1993ru}---where the thermal modification of the mass was computed through a one-loop perturbative correction of the pion self-energy---in this case, we solved a self-consistent set of equations, where the correction to the $D$-meson self-energy utilizes the full $T$-matrix at the vertices~\cite{Montana:2020vjg}. This $T$-matrix is now computed with a medium-modified retarded $G_{D\Phi}$ propagator,
\begin{align} 
  G_{D\Phi}(E,{\bf p};T) &=\int\frac{d^3q}{(2\pi)^3}\int d\omega\int d\omega'\frac{S_{D}(\omega,{\bf q};T)S_{\Phi}(\omega',{\bf p}-{\bf q};T)}{E-\omega-\omega'+i\varepsilon} \nn \\ & \times [1+f(\omega,T)+f(\omega',T)] \ , \label{eq:loopT2}
\end{align} 
where $f(\omega,T)$ is defined in Equation~(\ref{eq:BE}), and $S_D, S_\Phi$ denote the $D$-meson and light meson spectral functions, e.g.,
\begin{equation} \label{eq:specfunc}
  S_{D}(\omega,{\bf q};T)=-\frac{1}{\pi}{\rm Im\,}\mathcal{D}^R_{D}(\omega,{\bf q};T)=-\frac{1}{\pi}{\rm Im\,}\Bigg(\frac{1}{\omega^2-{\bf q}\,^2-m_{D}^2 (T=0)-\Pi^R_{D}(\omega,{\bf q};T)}\Bigg) \ ,
\end{equation}
which, in turn, depends on the retarded $D$-meson self-energy,
\begin{equation}\label{eq:selfE}
\Pi^R_{D}(\omega,{\bf q};T)=-\frac{1}{\pi} \int\frac{d^3q'}{(2\pi)^3}\int dE\frac{\omega}{\omega_\pi}\frac{f(E,T)-f(\omega_\pi,T)}{\omega^2-(\omega_\pi-E)^2+i\varepsilon} \ {\rm Im\,}T_{D\pi}(E,{\bf p};T) \ .
\end{equation}

These equations were numerically solved in Reference~\cite{Montana:2020lfi,Montana:2020vjg} in an iterative scheme. We considered temperatures up to $T=150$ MeV. In those works we checked that the effect of the pion thermal mass, as shown in the left panel (``SU(3) UChPT'' line) of Figure~\ref{fig:piChPT}, can be neglected in the thermal modification of the $D$-meson masses. Then, vacuum spectral functions for the light mesons were employed in order to simplify the calculation. 

The masses of the ground states are defined as the poles of the retarded $D$-meson propagator [${\cal D}^R_D$ in Equation~(\ref{eq:specfunc})] in the static limit ${\bf q}=0$, i.e., they are the solutions of \mbox{the equation,}
\be m_D^2(T)  = m_D^2(T=0) +\Pi_D^R(m_D(T),{\bf q}=0;T) \ , \ee
where $m_D (T=0)$ is the vacuum $D$-meson mass. For the masses of the dynamically-generated states, a simplified analysis was performed in~\cite{Montana:2020lfi,Montana:2020vjg} to render the problem tractable. Instead of analytically continuing the thermal scattering amplitudes to the full complex energy plane, we found the projection into the real energy axis of the different channels and extracted the masses and decay widths from standard spectral shapes. For the narrow states a Breit-Wigner-Fano form was employed, which takes into account the interference of the resonances with the background. For the the broad states in the $S=0$ channels---whose shapes are distorted due to the presence of several thresholds---we employed a three-coupled-channel Flatt\'e form, with the background subtracted. More details are given in Reference~\cite{Montana:2020vjg}.

\begin{figure}[H]
\begin{center}
\vspace{2pt}
\includegraphics[width=0.36\textwidth]{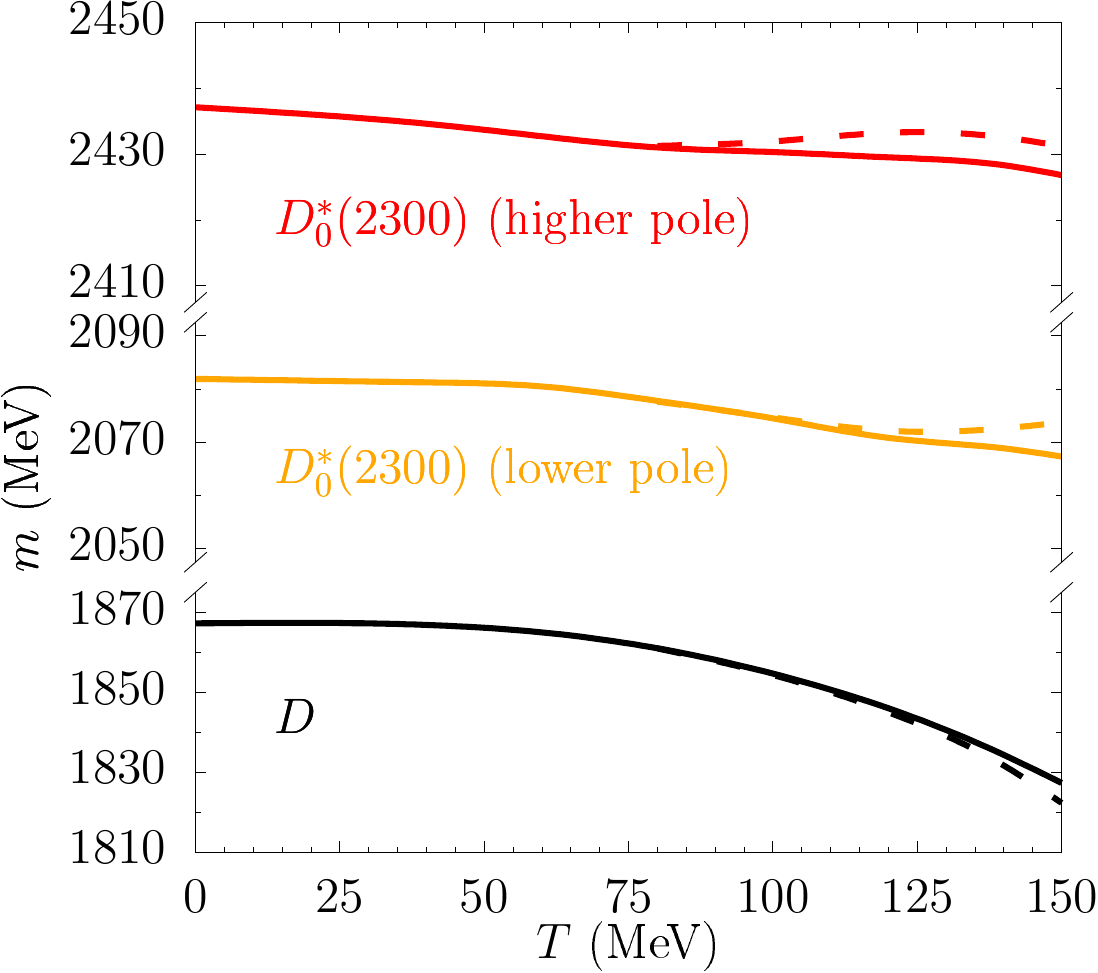}
\includegraphics[width=0.36\textwidth]{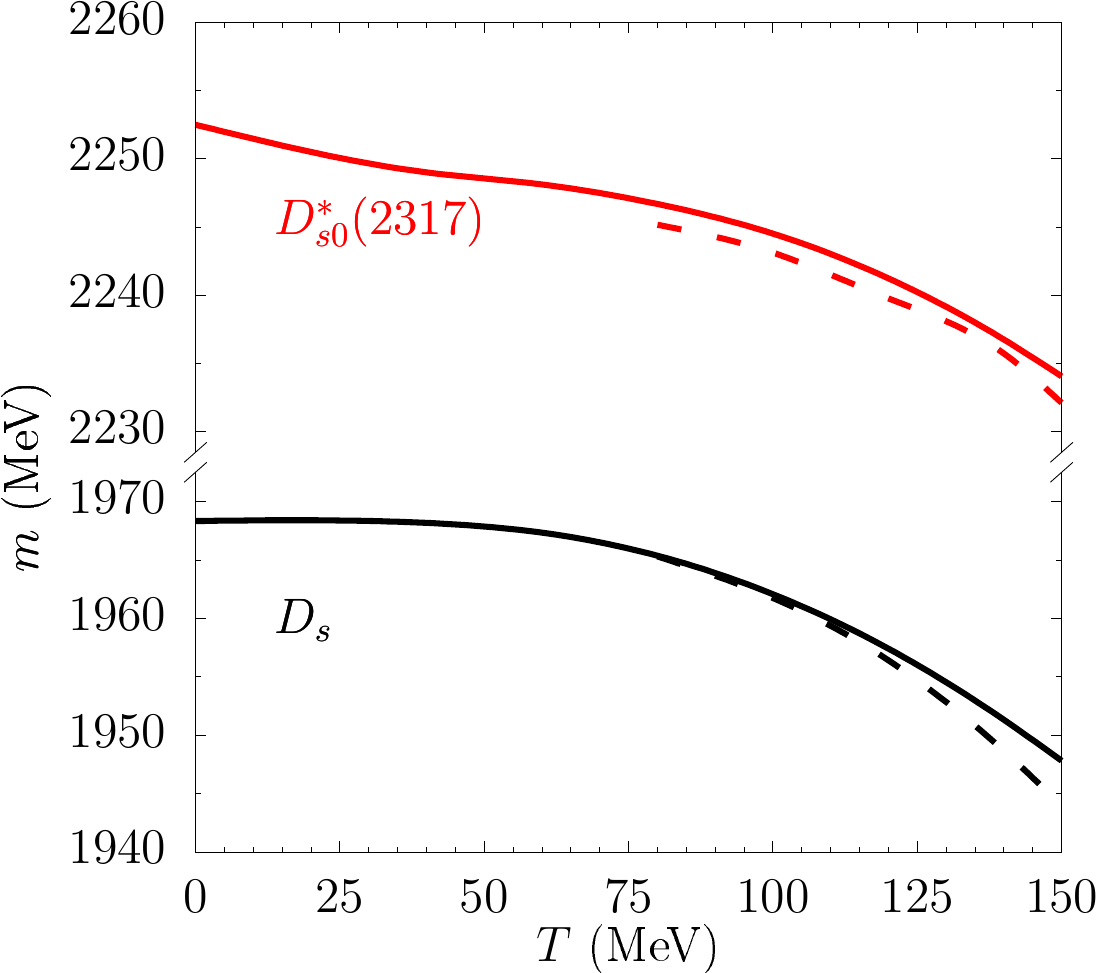}
\caption{\label{fig:thermalDJ0} Thermal masses of the $D$-meson chiral partners in the $J=0$ channel. (\textbf{Left panel}): $S=0$ channel with $D$ meson ground state and the $D_0^*(2300)$ resonance (double pole). (\textbf{Right panel}): $S=1$ channel with ground state $D_s$ meson and $D_{s0}^* (2317)$ bound state.}
\end{center}
\end{figure}

\begin{figure}[H]
\begin{center}
\includegraphics[width=0.36\textwidth]{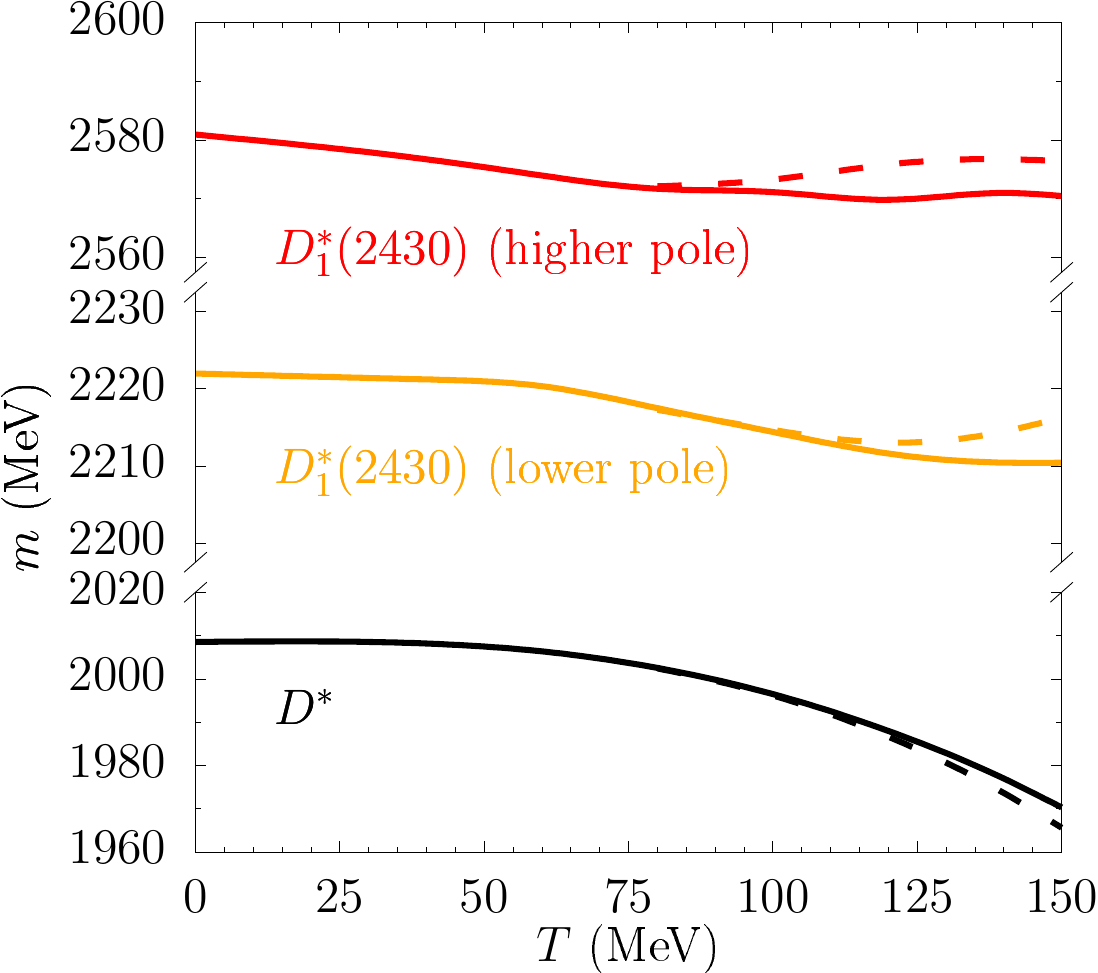}
\includegraphics[width=0.36\textwidth]{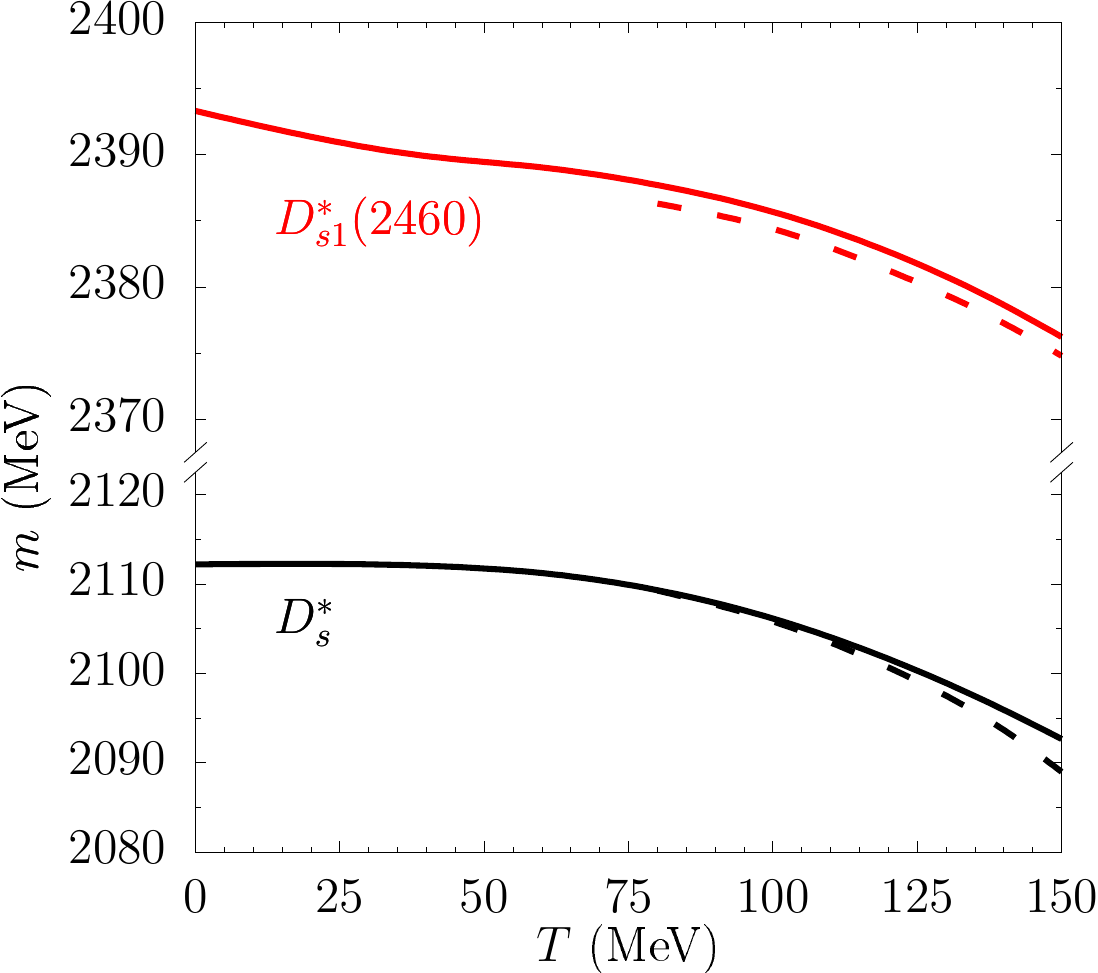}
\caption{\label{fig:thermalDJ1}   Thermal masses of the $D$-meson chiral partners in the $J=1$ channel. (\textbf{Left panel}): $S=0$ channel with $D^*$ meson ground state and the $D_1^*(2430)$ resonance (double pole). (\textbf{Right panel}): $S=1$ channel with ground state $D_s^*$ meson and $D_{s1}^* (2460)$ bound state. }
\end{center}
\end{figure}

I summarize the results of the thermal masses for the different $J=0$ states in Figure~\ref{fig:thermalDJ0}. The left and right panels show the results for the $S=0$ and the $S=1$ sectors, respectively. Ground states are represented in black lines. For completeness, I also present the result in the $J=1$ sector in Figure~\ref{fig:thermalDJ1}, which are very similar to those in $J=0$ sector thanks to the heavy-quark spin symmetry (apart from a generic mass shift of $\Delta m \simeq 140$ MeV due to the use of the physical vacuum masses for the ground states).

The different line style represents the set of light mesons used to dress the ground state mass in Equation~(\ref{eq:selfE}). The solid lines show the mass modification due to a thermal bath of pions, while the dashed lines contain the effect of pions, kaon and antikaons in a self-consistent way. As thermally suppressed, the effect of kaons states is very small and only appreciable at the highest temperatures.
\vspace{-6pt}

From Figures~\ref{fig:thermalDJ0} and \ref{fig:thermalDJ1} one observes a general dropping of the thermal masses when temperature increases. This goes in line with the $\pi$ and $\sigma$ masses presented in \mbox{Section~\ref{sec:chpt}}. However, the decrease is rather small in relative terms.  Nonstrange ground states $D$ and $D^*$ suffer a drop of $\simeq$2\% at $T=150$ MeV with respect to their vacuum masses. This reduction was also seen in the previous calculation~\cite{Fuchs:2004fh}. For the states with strangeness, $D_s$ and $D_s^*$, the drop is smaller ($\simeq$1\%). The mass reduction of the different resonances is even more limited, of the order of 10 MeV. Such effects are very small to conclude any evidence of mass degeneracy around $T=150$ MeV~\cite{Montana:2020lfi}. The reason is the large vacuum masses in comparison with the temperatures considered. What is important to notice is the fact that the chiral-partner degeneracy in the $S=0$ case involves not two but three different states. This is due to the double-pole structure of the positive parity state and consists of a new situation to explore.

\begin{figure}[H]
\begin{center}
\includegraphics[width=0.24\textwidth]{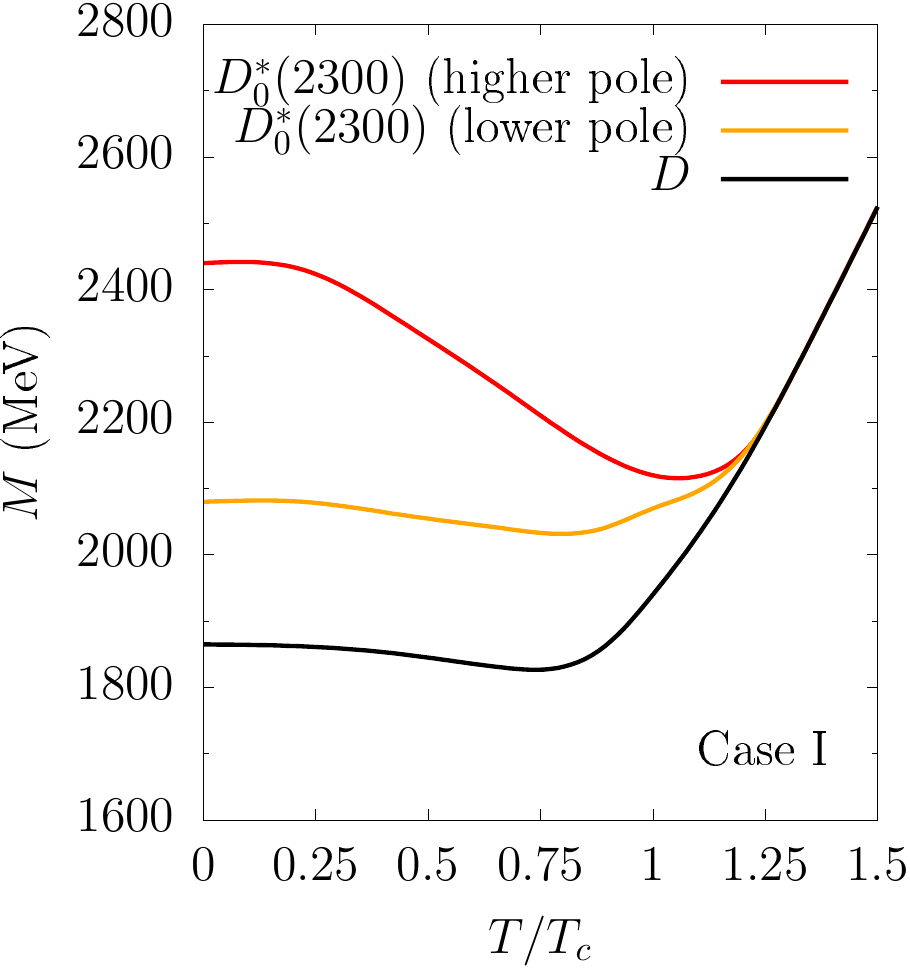}
\includegraphics[width=0.24\textwidth]{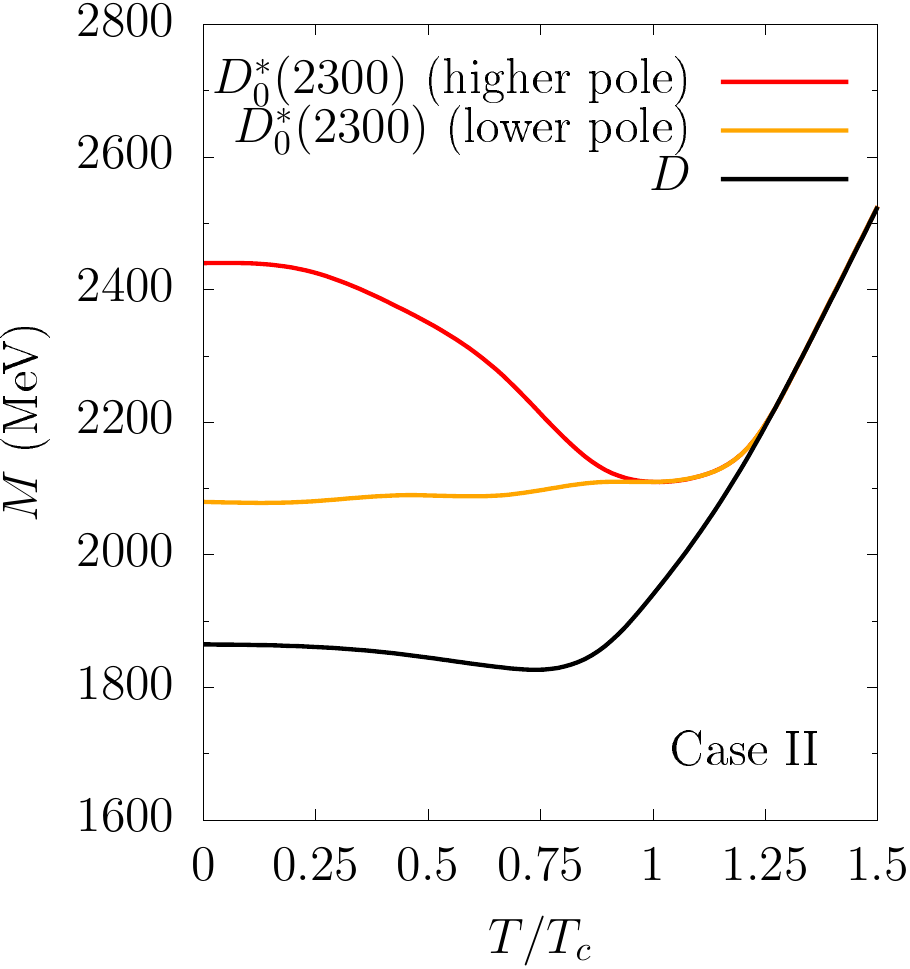}
\includegraphics[width=0.24\textwidth]{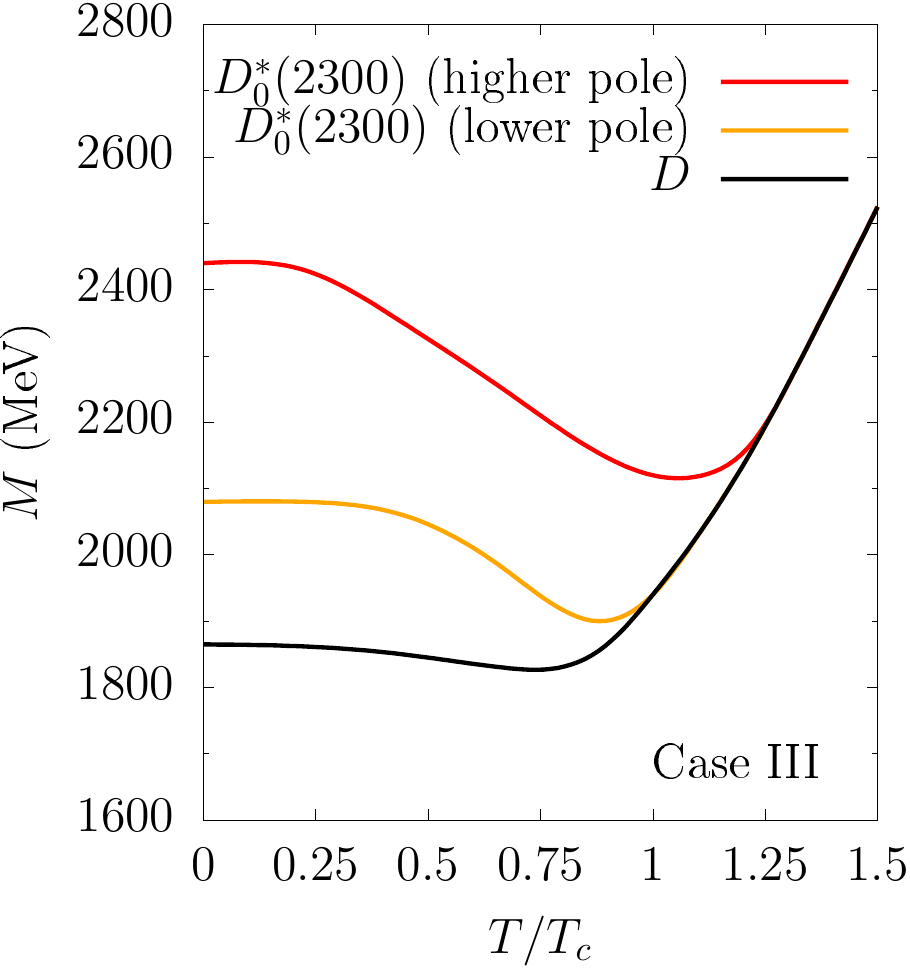}
\caption{\label{fig:cases} Three scenarios for the mass degeneracy of chiral companions, when the positive parity state present a double-pole structure. {(\textbf{Left panel})}: The three states become degenerate at approximately the same temperature above $T_c$. (\textbf{Middle panel}): The two-pole structure merges into a single pole before becoming degenerate with the chiral partner. (\textbf{Right panel}): The two poles of the chiral state become degenerate with the ground state at different temperatures above $T_c$.}
\end{center}
\end{figure}

As in ChPT, the EFT is limited below the chiral restoration temperature, and it is not possible to study the two-pole structure in the Weyl--Wigner phase. It is unclear if the double pole structure will merge into a single one before becoming degenerate with the ground state or if the three states will eventually get similar masses around the same temperature above $T_c$. Different scenarios for the chiral restoration are schematically shown in Figure~\ref{fig:cases}.

 In the first case the three states become degenerate around the same temperature above $T_c$. In the second case the two poles will merge before becoming degenerate with the ground state. In the third case the degeneracy occurs in a two-step process, first merging the lower pole with the ground state and later becoming degenerate with the upper pole. A more exotic scenario could in principle be considered, e.g., the true chiral companion of the ground state is only one of the poles, while the other never becomes degenerate. However, given that the physical state is a superposition of the two poles, it is difficult to imagine that such situation would correspond to the physical one.   

In the calculation of the $D$-meson masses of Reference~\cite{Montana:2020vjg} the thermal evolution of the pseudo-Goldstone bosons and the pion decay constant were neglected. For the first case we found that a reduced $\pi$ mass at $T=150$ MeV only contributed no more than 5 MeV to the final charm-meson masses (including the ground and the dynamically-generated states). In addition, no back reaction of the chiral condensate nor $f_\pi$ was taken into account in the calculation. However, the effect of the thermal dependence of the pion decay constant is expected to be more elucidating as it can be directly related to the chiral condensate via Gell-Mann-Oakes-Renner relation at finite temperature~\cite{Gasser:1986vb,Pisarski:1996mt,Toublan:1997rr,Weise:2001sg,Son:2002ci}. Notice that $f_\pi$ enters explicitly in the perturbative scattering amplitude, cf. Equation~(\ref{eq:potential}), and the effect of the chiral condensate was particularly crucial in the L$\sigma$M for the $\sigma$ state (see \mbox{Equation~(\ref{eq:msigmaT})}); and in the (P)NJL model for the dressing of the quark mass (see Equation~(\ref{eq:gap})). 

The implementation of $f_\pi(T)$---together with the thermal dependence of light meson properties---can shed some light on the preferred scenario among those described in \mbox{Figure~\ref{fig:cases}}. The ChPT sector of the EFT~(\ref{eq:lagrangianLO}), (\ref{eq:lagrangianNLO}) can provide a dependence of $f_\pi(T)$. From the works~\cite{Gasser:1986vb,Bochkarev:1995gi,Pisarski:1996mt} one would expect a reduction at lowest order (while at finite temperature one should distinguish between temporal and spatial pion decay constants, I will neglect such difference for the simple calculation presented here)
\be \frac{f_\pi(T)}{f_{\pi,0}} \simeq 1- \frac{T^2}{12f_{\pi,0}^2} \ , \ee
where $f_{\pi,0}=f_\pi(T=0)$.

Here I present a simplified calculation aiming for a preliminary implementation of the effects of the reduction of $f_\pi(T)$ on the two-pole structure of the $D_0^*(2300)$ state. I do not pursue a self-consistent framework and not even a systematic treatment using the imaginary-time formalism. Just to get a first taste I simply consider the vacuum calculation of the dynamically-generated poles by solving the $T$-matrix Equation~(\ref{eq:TmatrixD}), with an artificial modified value of $f_\pi$. Such a modified pion decay constant will take a fraction of the vacuum value $f_{\pi,0}=93$ MeV. I focus only on the scalar channel $J=0$ but including both $S=0$ and $S=1$ sectors.

\begin{figure}[H]
\begin{center}
\includegraphics[scale=0.62]{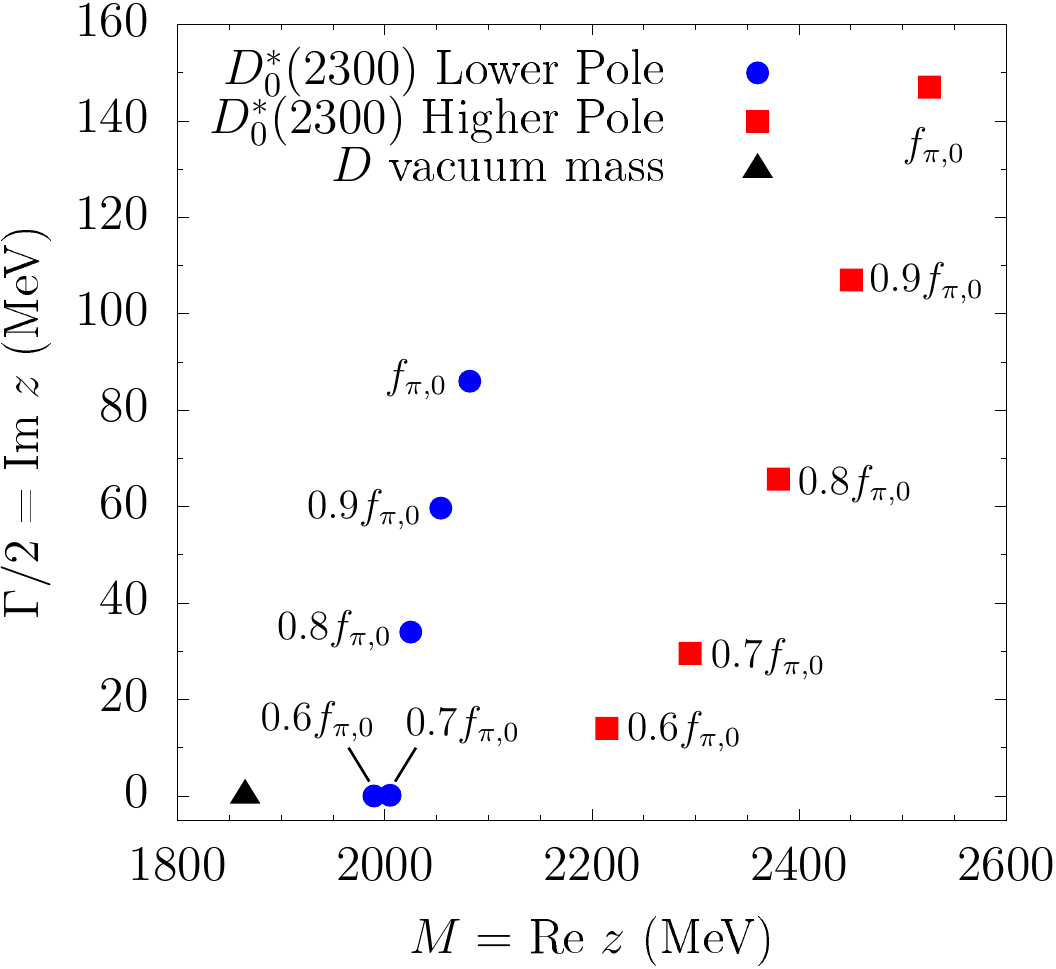}
\includegraphics[scale=0.62]{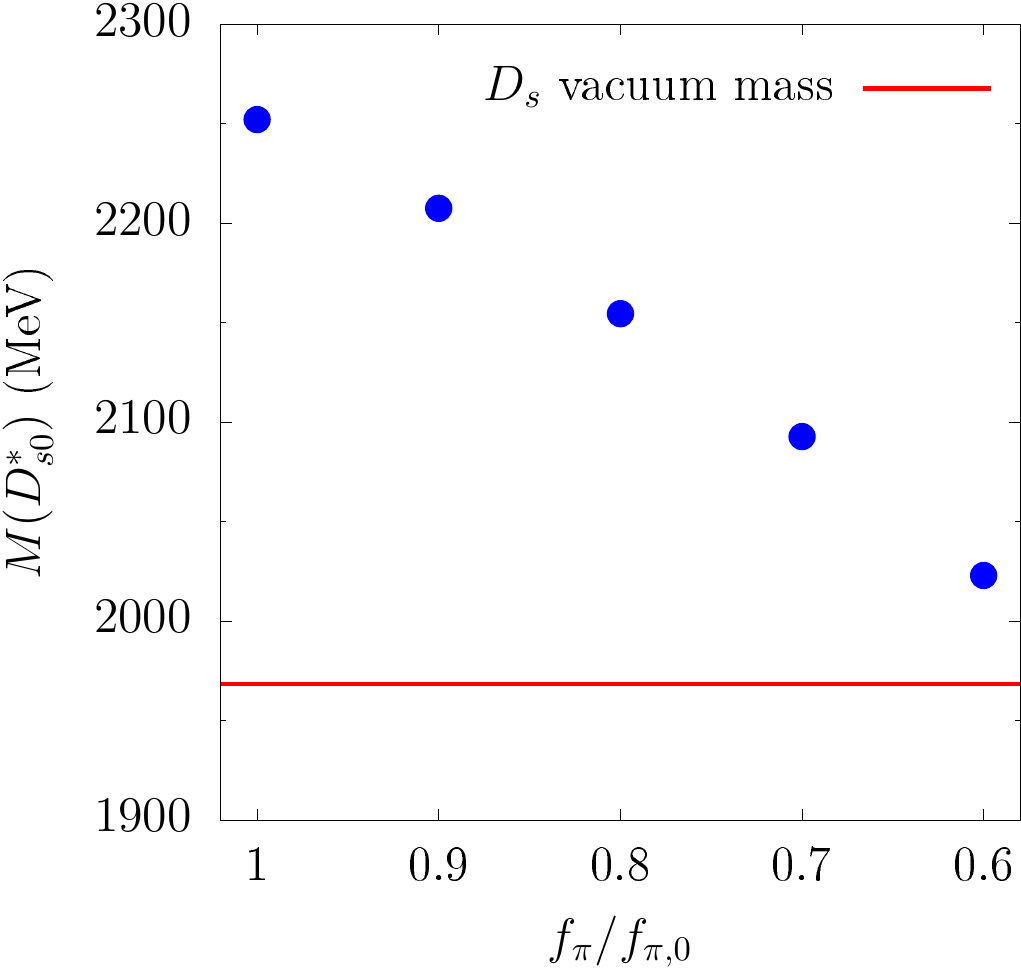}
\caption{Pole positions of the positive-parity charm states as functions of the pion decay constant, when it is reduced with respect to the vacuum value. (\textbf{Left panel}): $(J,S)=(0,0)$ channel where the two-pole state $D_0^*(2300)$ is generated. (\textbf{Right panel}): $(J,S)=(0,1)$ sector where the bound state $D_{s0}^* (2317)$ is obtained.~\label{fig:test}}
\end{center}
\end{figure}

The results are shown in Figure~\ref{fig:test}, where I allowed for a reduction of the pion decay constant up to 60 \% of its vacuum value. In the left panel I present the evolution of the two poles associated to the $D_0^* (2300)$ resonance as a function of $f_\pi$. Both poles get a sizable decrease of their real parts and a strong reduction of their decay widths. The higher pole remains in the original Riemann sheet (but monotonously approaching the real axis). The lower pole crosses from the $(+,-,-)$ Riemann sheet to the unphysical one $(+,+,+)$ thus becoming a bound state very close to $f_\pi=0.7f_{\pi,0}$. While the real parts of the pole positions are still far from the $D$-meson mass the approaching is evident and much more pronounced than the results in Figure~\ref{fig:thermalDJ0}. 

The situation in the $(J,S)=(0,1)$ channel is presented in the right panel of \mbox{Figure~\ref{fig:test}}. The dynamically-generated $D_{s0}^*$ remains bound for all values of $f_\pi$ considered and stays in the physical Riemann sheet. Then I only plot the real part of the pole position as a function of $f_{\pi}$ (in units of the vacuum pion decay constant). The positive-parity state clearly approaches the ground state mass, becoming rather close to it for $f_\pi=0.6 f_{\pi,0}$. This shows a potential mass degeneracy of chiral partners with a reduction of the effective pion decay constant.

It is even possible to account for the mass reduction of ground states---both charm and the light mesons---which I described in the previous sections. By repeating the exercise of varying $f_\pi$, but also implementing a $5\%$ reduction to the $D$ and $D_s$ masses (cf. black solid lines in Figure~\ref{fig:thermalDJ0}), and a $10\%$ reduction in all the light-meson masses (following the pion case in Figure~\ref{fig:piChPT}), I obtain a similar qualitative behaviors as in Figure~\ref{fig:test} but the real part of the pole positions of the $D_0^*(2300)$ and the pole of the $D_{s0}^*(2317)$ acquire an extra shift of $\sim$\ -100 MeV. In fact, both the lower pole of the $S=0$ channel and the bound state of $S=1$ sector, become very close to their respectively ground states. These analyses would favor the third scenario presented in Figure~\ref{fig:cases}, where the lower pole gets degenerate with the ground state before the higher pole does. 

One should keep in mind the very simplified nature of this study. I can only venture to conjecture that the two poles would become degenerate independently with the ground state, with the upper pole merging at a higher temperature. I hope this can motivate a more solid study in this direction, e.g., along the lines of~\cite{Sugiura:2019ane}, where mean-field effects of the quark condensate are incorporated in the determination of the vector $D^*$ mass. That could confirm whether the two-pole structure will not collapse into a single one before becoming degenerate with the ground state.

I conclude this section by mentioning that the scenario of three chiral siblings is not exceptional. Due to heavy-quark spin-flavor symmetry, analogous situations are expected for the nonstrange sectors of the $D^*$, $B$ and $B^*$ mesons (see~\cite{Meissner:2020khl} and references therein). On the other hand, the two-pole structure of the $\Lambda(1405)$ baryon~\cite{Oller:2000fj,GarciaRecio:2002td,Jido:2003cb} is well established. This state can be identified with the chiral partner of the $\Lambda$ ground state. Other meson and baryon candidates with a possible double-pole structure are mentioned in Reference~\cite{Meissner:2020khl}. To the best of my knowledge no other study on chiral symmetry restoration in the context of two-pole structures have been performed apart from~\cite{Montana:2020lfi}. For example, it would be interesting to address the thermal evolution of the two-pole structure of the $\Lambda(1405)$ at finite temperature, together with the thermal modification of the $\Lambda$ baryon.

\section{Conclusions}

  In this work I have reviewed some results concerning how chiral partners of opposed parity become degenerate in mass at high temperatures following the restoration of chiral symmetry. I have characterized the different patterns depending on the nature of the chiral states, namely, they are part of the fundamental degrees of freedom of the EFT or dynamically-generated states emerging from the two-body dynamics.
  
  In particular I have considered the scenarios where two chiral partners are degrees of freedom of the Lagrangian; both are emergent collective states; and the mixed case, where the positive parity state is generated by the attractive interactions of the chiral companion. These cases have been represented respectively by the L$\sigma$M, the (P)NJL model and the ChPT. For the latter case---even when incorporating the correct chiral constraints in its construction---the effective theory is limited to temperatures below the chiral restoration one. Therefore, a clear conclusion for a full mass degeneracy cannot be drawn. However, evidences of a relative approach around $T=125$ MeV were found when employing a unitarized version of the ChPT.
  
  Finally, I have addressed the case where the chiral multiplet is represented by three different states. This situation occurs in the heavy-light meson sector, where the chiral companion of the $D$ meson is the $D_0^*(2300)$ state, which is a combination of two poles in the complex-energy plane. A similar situation occurs in the $J=1$ sector between the $D^*$ meson and the double-pole structure of the $D_1^*(2430)$ state. In the case of ChPT definite results cannot be obtained, being that the temperature dependence is very weak for the mass of these heavy states. Nevertheless, I argued that a coupling with the chiral condensate might bring a more solid conclusion in this respect. I have presented a preliminary study of the thermal masses of the positive-parity states by artificially decreasing the value of $f_\pi$. I found a substantial $T$ dependence pointing to a decrease of the mass difference between chiral states. While this was a mere exploratory study, it seems to indicate that a potential degeneracy of chiral siblings at high temperatures might not occur for the three states at once but sequentially. A rigorous description of how chiral symmetry restoration is realized with three chiral companions (ground state and double-pole structure) is still an open issue. I hope future investigations can shed light on this intriguing situation.

\vspace{6pt} 



\funding{This research was funded by the Deutsche Forschungsgemeinschaft (German Research Foundation) grant numbers 411563442 (Hot Heavy Mesons) and 315477589---TRR 211 (Strong-interaction matter under extreme conditions).
}

\institutionalreview{Not applicable}

\informedconsent{Not applicable}

\dataavailability{The data supporting the presented results are available in their corresponding references, as detailed in the main text. Data (in tabular form) related to any of the shown figures are available on request from the corresponding author.
}

\acknowledgments{The author would like to warmly thank all his current and past collaborators, with whom he has obtained part of the results presented in this work.}

\conflictsofinterest{The author declares no conflict of interest.} 

\abbreviations{Abbreviations}{
The following abbreviations are used in this manuscript:\\

\noindent 
\begin{tabular}{@{}ll}
ChPT & Chiral perturbation theory\\
EFT & Effective field theory\\
IAM & Inverse amplitude method \\
LO & Leading order \\
L$\sigma$M & Linear sigma model\\
NJL & Nambu--Jona-Lasinio model\\
NLO & Next-to-leading order \\
PNJL & Polyakov--Nambu--Jona-Lasinio model \\
QCD & Quantum chromodynamics \\
RS & Riemann sheet \\
SSB & Spontaneous symmetry breaking\\
UV & Ultraviolet \\
VEV & Vacuum expectation value\\
\end{tabular}}

%

\end{paracol}
\reftitle{References}

\end{document}